\documentstyle[12pt,epsf]{article}

\begin{document}

\baselineskip=18.8pt plus 0.2pt minus 0.1pt

\makeatletter

\renewcommand{\thefootnote}{\fnsymbol{footnote}}
\newcommand{\beq}{\begin{equation}}
\newcommand{\eeq}{\end{equation}}
\newcommand{\bea}{\begin{eqnarray}}
\newcommand{\eea}{\end{eqnarray}}
\newcommand{\nn}{\nonumber}
\newcommand{\hs}[1]{\hspace{#1}}
\newcommand{\vs}[1]{\vspace{#1}}
\newcommand{\Half}{\frac{1}{2}}
\newcommand{\p}{\partial}
\newcommand{\ol}{\overline}
\newcommand{\wt}[1]{\widetilde{#1}}
\newcommand{\ap}{\alpha'}
\newcommand{\bra}[1]{\left\langle  #1 \right\vert }
\newcommand{\ket}[1]{\left\vert #1 \right\rangle }
\newcommand{\vev}[1]{\left\langle  #1 \right\rangle }
\newcommand{\vac}{\ket{0}}

\newcommand{\ul}[1]{\underline{#1}}

\makeatother

\begin{titlepage}
\title{
\hfill\parbox{4cm}
{\normalsize KIAS-P01026\\{\tt hep-th/0105246}}\\
\vspace{1cm}
Tachyon Lump Solutions of Bosonic D-branes on SU(2) Group Manifolds in 
Cubic String Field Theory
}
\author{Yoji Michishita
\thanks{
{\tt michishi@kias.re.kr}
}
\\[7pt]
{\it School of Physics, Korea Institute for Advanced Study}\\
{\it 207-43,Cheongryangri, Dongdaemun, Seoul, 130-012, Korea}
}

\date{\normalsize May, 2001}
\maketitle
\thispagestyle{empty}

\begin{abstract}
\normalsize
We construct tachyon lump solutions of bosonic D2-branes on SU(2) 
group manifolds by level truncation approximation in cubic string field 
theory, which are regarded as D0-branes.
The energies for these solutions show good agreement with 
the expected values.
\end{abstract}

\vspace{3cm}



\end{titlepage}

\section{Introduction}

Tachyon condensation on D-branes has been studied extensively.
It is conjectured for bosonic D-branes that the minimum of the
tachyon potential represents closed string vacuum \cite{s1}, and
tachyon lump solutions represent lower dimensional
branes \cite{s1,rscclmpt}.

String field theories are powerful tools for checking these 
conjectures. In many cases one of the following two types of string
field theories has been used : cubic
string field theory (CSFT) \cite{w1} and boundary string field theory
(BSFT) \cite{boundary}. By level truncation approximation 
in CSFT we can calculate numerically the value of the potential
at the minimum \cite{sz,bpot}, construct tachyon lump 
solutions \cite{lumps}, and confirm the absence of physical excitation 
on the vacuum \cite{nophys}. Furthermore, the form of the 
action around the closed string vacuum (vacuum SFT) is 
conjectured in \cite{vcsft} and its properties are investigated in
\cite{vlumpsetc,rsz}.
On the
other hand, by using BSFT we can get the exact tachyon potential and lump 
solutions \cite{bsft}. (For related works and the case of superstring see 
\cite{super,hsms,superb}. For a review see \cite{o1}.) At
present the relation between these two string field 
theories is unclear. Therefore it is worth investigating the tachyon
condensation by both of them. In this paper we use CSFT and level 
truncation approximation.

In CSFT it is proven that the tachyon potential can be determined
irrelevantly to the choice of closed string background \cite{s2}. 
The crucial point for the proof of this fact is that we can 
drop all the fields except the unit operator and its
descendants. However, for lump solutions we must give nonzero vev to
nontrivial primary operators and their descendants, and therefore the
results depend on the choice of closed string background.
The solutions constructed in \cite{lumps} are those on 
flat background. 

For further understanding of tachyon condensation, it is desirable to
consider lump solutions on curved backgrounds. In this paper we
give such an example: lump solutions on SU(2) 
group manifold. SU(2) group manifold is an example of curved 
background on which the worldsheet CFT is known exactly i.e. WZW model.
This manifold is also known as the example which has the noncommutative 
structure \cite{ars1,ars2}. The analyses of tachyon condensation by 
the worldsheet CFT
and the effective action is given in \cite{hnt} 

D-branes which extend to two dimensions in SU(2) (we call them
D2-branes) can be regarded as bound states of D-branes which sit
at a point in SU(2) (we call them D0-branes) \cite{bds,ars2}, 
and D0-branes are conjectured to be able to vanish into the vacuum by
tachyon condensation. Therefore one can imagine the process that some
of the D0-branes forming the D2-brane vanish and the other D0-branes
remain. Hence it is natural to conjecture that there exist lump
solutions on D2-branes which represent some D0-branes. 

We construct such lump solutions in CSFT in level [1,3], [2,6] and [5/2,5]
approximation, where [$N$,$M$] means that we consider only the fields with
the level (i.e. conformal dimension$+1$) $<N$ and retain only 
the terms with the sum of the levels $<M$. (On the other hand the 
notation ($N$,$M$) used in many literature means only the fields 
with the level $\leq N$ and only the terms 
with the sum of the levels $\leq M$.)
We consider the case where the D2-brane is the bound state of 2 and 3 
D0-branes.

This paper is organized as follows. In section \ref{sec2}, we 
summarize the
properties of D-branes on SU(2) group manifold and the cubic string
field theory on it, and explain how to calculate the terms
in the action of CSFT. In section \ref{sec3}, we construct 
lump solutions and compute their energy. Section \ref{sec4} 
contains conclusions and discussions.

In the following $\ap$ is taken to be 1.

\section{D-branes and cubic string field theory on SU(2) group manifolds
\label{sec2}}
In this section we summarize the known results about D-branes of
bosonic string theory on SU(2) group manifolds and fix the notations.
Then we explain how to calculate the terms in the action of cubic string
field theory on these manifolds.
\subsection{The worldsheet CFT}
SU(2) group manifold ($\simeq S^3$) can be, for example, parametrized
as follows.
\beq
g=\left(\begin{array}{cc}
x_4-ix_2 & -x_1-ix_3 \\ x_1-ix_3 & x_4+ix_2, \\
\end{array}\right)
\in SU(2),
\eeq
\bea
x_1 & = & \sin\psi\sin\theta\cos\phi, \nn\\
x_2 & = & \sin\psi\sin\theta\sin\phi, \nn\\
x_3 & = & \sin\psi\cos\theta, \nn\\
x_4 & = & \cos\psi,
\eea
where $\theta,\phi$ and $\psi$ are spherical coordinates of $S^3$.
Let the radius of $S^3$ be $\sqrt{k\ap}=\sqrt{k}$.
The worldsheet CFT on SU(2) group manifold, i.e. SU(2) WZW model, has
the left moving current $J^a(z)$ and the right moving current
$\bar{J}^a(\bar{z})$. Their mode expansions are
$J^a(z)=\sum_{n}z^{-n-1}J_n^a$ and similar one for
$\bar{J}^a(\bar{z})$. The commutation relation of these modes is 
\beq
[J_n^a,J_m^b]=if^{abc}J_{n+m}^c+kn\delta^{ab}\delta_{n,-m}.
\label{commutation}
\eeq
($f^{abc}=\epsilon^{abc}$ for SU(2))
The radius $k$ becomes the level of the SU(2) current
algebra. Therefore $k$ is quantized to be positive integer valued.   
The central charge of this CFT is $\frac{3k}{k+2}$.
$J_0^a$ form ordinary SU(2) algebra and primary
fields belong to the representations of this algebra.
We denote the primary field in spin $j$ representation by $\Phi_j^m$. 
Then,
\beq
J_0^a \ket{\Phi_j^m} = 
 \ket{\Phi_j^{m'}} (T^{a}_{(j)})^{\;\; m}_{m'},\quad
J_{n>0}^a \ket{\Phi_j^m} = 0,
\eeq
where $(T^{a}_{(j)})^{\;\; m}_{m'}$ is the spin $j$ representation matrix.
The conformal dimension of $\Phi_j^m$ is $h_j\equiv \frac{j(j+1)}{k+2}$.
We take the diagonal modular invariant as the closed string 
1-loop partition function.

When we consider open strings we must impose some
boundary condition which relates $J^a(z)$ to $\bar{J}^a(\bar{z})$.
Henceforth we consider the following condition.
\beq
J^a(z) = \bar{J}^a(\bar{z}) \mbox{ on the boundary} ,
\eeq
where $z$ is a coordinate of the upper half plane and we take the real
axis as the boundary. Boundary states for this boundary condition are
\cite{c,iio}
\beq
\ket{J} = \sum_j
\frac{S_J^j}{\sqrt{S_0^j}}{\vert j \rangle\rangle},
\eeq
where ${\vert j \rangle\rangle}$ is the Ishibashi state constructed on the
primary field $\Phi_j^m$ and $S_J^j$ is the modular
transformation matrix:
\beq
S_J^j = \sqrt{\frac{2}{k+2}}\sin{\frac{(2J+1)(2j+1)\pi}{k+2}}.
\eeq
$J$ takes the values $0,\Half,1,\cdots,\frac{k}{2}$. The worldvolume of 
the boundary state labelled by $J$ is $\psi={\rm const.}=
\frac{(2J+1)\pi}{k+2}$,
which is an $S^2$ parametrized by $\theta$ and $\phi$ \cite{as,fffs}. 
If $J=0$ or
$\frac{k}{2}$ ,the $S^2$ shrinks to a point (at $k\rightarrow\infty$). 
Thus $\ket{J}$ with
$J=0,\frac{k}{2}$ represent D0-branes and others are D2-branes.
The D2-brane with boundary state $\ket{J}$ can be regarded as the bound
state of $(2J+1)$ D0-branes \cite{bds,ars2}.

The spectrum of primary fields on the brane with boundary state
$\ket{J}$ is 
\beq
\Phi_j^m,\quad j=0,1,2,\cdots,{\rm min}(2J,k-2J).
\eeq
Note that $j$ takes only integers. Their OPE is known as follows 
\cite{r,ars1}.
\beq
\Phi_{j_1}^{m_1}(x)\Phi_{j_1}^{m_1}(y) =
 (x-y)^{h_{j_3}-h_{j_1}-h_{j_2}}
 \vev{j_1,m_1,j_2,m_2 | j_1,j_2,j_3,m_3} c_{j_1 j_2}^{j_3}(k,J)
 \Phi_{j_3}^{m_3}(y) + (\mbox{descendants}),
\label{ope}
\eeq
where $\vev{j_1,m_1,j_2,m_2 | j_1,j_2,j_3,m_3}$ is the Clebsch-Gordan
coefficient, and $c_{j_1 j_2}^{j_3}(k,J)$ is written by quantum 6-j
symbol and given in the appendix \ref{appa}. 
Three point correlation functions are 
\beq
\bra{\Phi_{j_1}^{m_1}}\Phi_{j_2}^{m_2}(1)\ket{\Phi_{j_3}^{m_3}} =
  C\left[\begin{array}{ccc}
  m_1 & m_2 & m_3 \\
  j_1 & j_2 & j_3 \\
   \end{array}\right],
\label{3point}
\eeq
and
\beq
C\left[\begin{array}{ccc}
  m_1 & m_2 & m_3 \\
  j_1 & j_2 & j_3 \\
   \end{array}\right]
 =\frac{(-1)^{m_3}}{\sqrt{(2j_3+1)[2j_3+1]}}
 \vev{j_1,m_1,j_2,m_2 | j_1,j_2,j_3,-m_3} c_{j_1 j_2}^{j_3}(k,J).
\eeq
The definition of $[2j_3+1]$ is given in the appendix \ref{appa}.
$C\left[\begin{array}{ccc}
  m_1 & m_2 & m_3 \\
  j_1 & j_2 & j_3 \\
   \end{array}\right]$
is symmetric under cyclic permutations of (1,2,3), as should be for
the three point functions, and satisfies the following relations.
\bea
C\left[\begin{array}{ccc}
  m_1 & m_2 & m_3 \\
  j_1 & j_2 & j_3 \\
   \end{array}\right]
& = & 0 \quad{\rm if}\; m_1+m_2+m_3\neq 0,\\
C\left[\begin{array}{ccc}
  m_1 & m_2 & m_3 \\
  j_1 & j_2 & j_3 \\
   \end{array}\right]
& = & (-1)^{j_1+j_2+j_3}C\left[\begin{array}{ccc}
  m_2 & m_1 & m_3 \\
  j_2 & j_1 & j_3 \\
   \end{array}\right].
\label{ccyc}
\eea

\subsection{Cubic string field theory on SU(2)}
Next we explain the cubic string field theory and how to calculate the 
terms in the action.
We will consider the following background.
\beq
X\times Y \times {\cal M} \times {\rm SU(2)},
\eeq
where $X$ is the time direction which is taken to be flat, and $Y$ is 
a spatial direction, which is also taken to be flat, and the boundary 
condition for open strings along this direction is taken to be 
Dirichlet. ${\cal M}$ is an arbitrary background with the central 
charge $24-\frac{3k}{k+2}$, appropriate for cancelling the total 
central charge.
We denote the Virasoro operator of the matter part of 
$X\times Y \times {\cal M} \times {\rm SU(2)}$ and 
$X\times Y \times {\cal M}$ by $L^{(m)}_{n}$ and $L'_{n}$ respectively,
and the central charge of $X\times Y \times {\cal M}$ by 
$c'=26-\frac{3k}{k+2}$

The action for string field $\Psi$ on general CFT background is 
constructed in \cite{s3}:
\beq
S=-\frac{1}{g_o^2}\left(\Half\vev{\Psi ,Q_B\Psi}
+\frac{1}{3}\vev{\Psi,\Psi,\Psi}\right).
\eeq
We follow the notation of \cite{rz}.
$Q_B$ is the BRST charge:
\bea
Q_B & = & \oint\frac{dz}{2\pi i}:c(z)
 \left(T^{(m)}(z)+\Half T^{{gh}}(z)\right): \nn\\
 & = & \sum_{n=-\infty}^{\infty}c_{-n}L_{n}^{(m)}
 + c_0 \Bigl[ \sum_{n\geq 1}n(c_{-n}b_n+b_{-n}c_n)-1\Bigr]
 - 2b_0 \sum_{n\geq 1}nc_{-n}c_n \nn\\
& & - \sum_{n\geq 1,m\geq 1} 
 \Bigl[ m(b_{-n-m}c_nc_m-c_{-n}c_{-m}b_{n+m}) \nn \\
& & +(n+2m)(b_{-n}c_{-m}c_{m+n}+c_{-m-n}c_mb_n)\Bigr].
\eea
Thanks to the presence of $Y$ we can relate $g_o$ to the mass of
D-brane $M$ \cite{s2}:
\beq
\frac{1}{g_o^2}= 2\pi^2 M.
\eeq
We can calculate $M$ by computing the cylinder amplitude and factorizing 
the contribution of gravitons, or computing the disk partition 
function \cite{ars2,bds}. The mass $M_{J}$ of the
brane represented by the boundary state $\ket{J}$ is
\beq
M_{J}\propto S_J^0=\sqrt{\frac{2}{k+2}}\sin{\frac{(2J+1)\pi}{k+2}}.
\eeq
The proportionality coefficient is determined from the information
of ${\cal M}$. We do not have to know it for our purpose.

The string field $\Psi$ can be expanded by primary operators and their 
descendants. We do not have to consider all of them for seeking lump
solutions on D-branes which depends only on the SU(2) directions. The 
components which we have to consider are those made by acting with the
oscillators 
\beq
\{J^a_{-1},J^a_{-2},\cdots; L'_{-2},L'_{-3},\cdots; 
c_{0},c_{-1}\cdots; ,b_{-1},b_{-2},\cdots\}
\eeq
on $\ket{j,m}$, where $\ket{j,m}=\Phi^m_j(0)c_1\vac$ and $\vac$ is the
SL(2,R) invariant vacuum.
It can be easily shown that the set of the other components do
not have linear terms in the action and therefore their equations of
motion are satisfied if they vanish. Hence they can be taken to be
zero from beginning. 
Then $\Psi$ can be expanded as follows.
\bea
 \Psi & = & \ket{j,m} t_m^j \nn\\
 & & + J_{-1}^a\ket{j,m} u_m^{aj} \nn\\
 & & + J_{-2}^a\ket{j,m} v_m^{aj}
  + J_{-1}^aJ_{-1}^b\ket{j,m} v_m^{abj}
  + b_{-1}c_{-1}\ket{j,m}\beta_m^{aj}
  + L'_{-2}\ket{j,m} w_m^j \nn\\
 & & + \cdots \nn\\
 & & +c_0 \Bigl[ b_{-1}\ket{j,m}\wt{t}_m^j \nn\\
 & & + J_{-1}^ab_{-1}\ket{j,m} \wt{u}_m^{aj}
  + b_{-2}\ket{j,m} \wt{u}_m^j \nn\\
 & & + \cdots\Bigr],
\label{stfield}
\eea
Here the repeated indices are summed over.
The terms containing $c_0$ are removed if we take the Siegel
gauge $b_0\Phi=0$.
$\ket{j,m}$ and its descendants correspond to the spherical harmonic
$Y_j^m$ \cite{ars2,fffs}.
Since $(Y_j^m)^*=(-1)^mY_j^{-m}$, we impose the following reality 
conditions.
\bea
& & (t_m^j)^*=(-1)^mt_{-m}^j,\quad 
(u_m^{aj})^*=(-1)^mu_{-m}^{aj},\quad 
(v_m^{abj})^*=(-1)^mv_{-m}^{abj},\nn\\
& & (\beta_m^j)^*=(-1)^m\beta_{-m}^j,\quad
(w_m^j)^*=(-1)^mw_{-m}^j,\quad
(\wt{u}_m^{aj})^*=(-1)^m\wt{u}_{-m}^{aj},\nn\\
& & (v_m^{aj})^*=-(-1)^mv_{-m}^{aj},\quad
(\wt{t}_m^j)^*=-(-1)^m\wt{t}_{-m}^j,\quad
(\wt{u}_m^j)^*=-(-1)^m\wt{u}_{-m}^j,\quad\mbox{etc.}
\label{reality}
\eea
Additional sign factor is needed for $v_m^{aj}$, $\wt{t}_m^j$ and 
$\wt{u}_m^j$ because their BPZ conjugations are equal to minus 
the hermitian conjugations.

We can reduce the number of components further.
The components with the eigenvalues of $J_0^3\neq 0$ has no linear
terms because of global SU(2) symmetry generated by
$J^a_0$. Therefore we can satisfy the equations of motion by putting
them to be zero. We consider only the components with $J_0^3=0$. 
Then we can restrict the components in (\ref{stfield}) 
in the following form.
\bea
& & t_m^j=t^j \delta_{0m}, \quad 
\beta_m^j=\beta^j \delta_{0m}, \quad 
w_m^j=w^j \delta_{0m}, \quad 
\wt{t}_m^j=i\wt{t}^j \delta_{0m}, \quad
\wt{u}_m^j=i\wt{u}^j \delta_{0m}, \nn\\
& & u_m^{aj}=\eta^{j+j'+1}U^{m'a}\vev{1,m',j,m|1jj'0}u^{jj'}, \nn\\
& & v_m^{aj}=\eta^{j+j'}U^{m'a}\vev{1,m',j,m|1jj'0}v^{jj'}, \nn\\
& & \wt{u}_m^{aj}=\eta^{j+j'+1}U^{m'a}\vev{1,m',j,m|1jj'0}\wt{u}^{jj'}, \nn\\
& & v_m^{abj}=\eta^{j+j''}U^{m_3b}\vev{1,m_3,j',m_2|1j'j''0}
 U^{m_1a}\vev{1,m_1,j,m|1jj'm_2}v^{jj'j''},
\eea
where $\eta^j=1\mbox{ (for }j=\mbox{even)},i\mbox{ (for }j=\mbox{odd)}$
and $U^{ma}$ is the following matrix which convert the
standard spin 1 index $m$ to the adjoint index $a$.
\beq
U^{ma}=\left(\begin{array}{ccc}
-\frac{1}{\sqrt{2}} & -\frac{i}{\sqrt{2}} & 0 \\
0 & 0 & 1 \\
\frac{1}{\sqrt{2}} & -\frac{i}{\sqrt{2}} & 0
\end{array}\right).
\eeq
The reality condition for these components are determined from 
(\ref{reality}). All of them are real. i.e.
\bea
& & (t^j)^*=t^j, \quad 
(\beta^j)^*=\beta^j, \quad 
(w^j)^*=w^j, \quad 
(\wt{t}^j)^*=\wt{t}^j, \quad
(\wt{u}^j)^*=\wt{u}^j, \nn\\
& & (u^{jj'})^*=u^{jj'},\quad (v^{jj'})^*=v^{jj'},\quad
(\wt{u}^{jj'})^*=\wt{u}^{jj'},\quad (v^{jj'j''})^*=v^{jj'j''}.
\eea

Now let us explain the calculation of the quadratic terms in the action.
First, we get 2-point functions of two SU(2) primary
operators by putting $j_2=m_2=0$ in eq.(\ref{3point}) ($\Phi^0_0=1$).
\bea
\vev{\Phi_{j}^{m}|\Phi_{j'}^{m'}} & = &
  C\left[\begin{array}{ccc}
  m & 0 & m' \\
  j & 0 & j' \\
   \end{array}\right] \nn\\
 & = & \frac{(-1)^m}{\sqrt{(2j+1)\ [2j+1]}}\delta_{j,j'}\delta_{m,-m'}.
\eea
The BPZ conjugation of $J_{-n_1}^{a_1}J_{-n_2}^{a_2}\cdots\ket{\Phi_j^m}$
is $\bra{\Phi_j^m}\cdots [(-1)^{n_2+1}J_{n_2}^{a_2}]
[(-1)^{n_1+1}J_{n_1}^{a_1}]$. Using this and the commutation relation
(\ref{commutation}),
2-point functions of descendants can be reduced to those of primaries.

Let $\Psi_1=\phi_1 G_1\vac$ and $\Psi_2=\phi_2 G_2\vac$ 
, where $G_i$ are ghost parts, and $\phi_i$ are SU(2) parts.
Then
\beq
\vev{\Psi_1,Q_B\Psi_2} = \oint\frac{dz}{2\pi i} \left[ 
\vev{G_1|c(z)|G_2}\vev{\phi_1|T^{(m)}(z)|\phi_2}
+\Half\vev{G_1|:c(z)T^{(gh)}(z):|G_2}\vev{\phi_1|\phi_2} \right].
\label{quadratic}
\eeq
Since $\oint\frac{dz}{2\pi i}:c(z)T^{(gh)}(z):$ does not change
descendant level and 
$\vev{\phi_1|\phi_2}$ is nonzero only when $\phi_1$ and $\phi_2$ have
the same descendant level, the second term of (\ref{quadratic}) is
nonzero only when $\Psi_1$ and $\Psi_2$ have the same SU(2) and ghost
descendant level. 
$\vev{\phi_1|T^{(m)}(z)|\phi_2}$ is nonzero only when
$T^{(m)}(z)\ket{\phi_2}$ and $\ket{\phi_1}$ have the same descendant
level. Therefore
$\vev{\phi_1|T^{(m)}(z)|\phi_2}
=z^{N_1-N_2-2}\vev{\phi_1|L^{(m)}_{N_2-N_1}|\phi_2}$,
where $N_i$ are the descendant level of $\Psi_i$. Then the first term
of (\ref{quadratic}) is equal to
\beq
\vev{G_1|c_{N_1-N_2}|G_2}\vev{\phi_1|L^{(m)}_{N_2-N_1}|\phi_2}.
\eeq
This is nonzero only when $\Psi_1$ and $\Psi_2$ have the same total
descendant level. Thus we can reduce the number of terms needed to
calculate. Remaining terms can be calculated by using commutation
relations of $L^{(m)}_n,J_n^a,c_n$ and $b_n$.

(\ref{quadratic}) becomes particularly simple when
$\ket{G_1}$ and $\ket{G_2}$ do not contain $c_0$ i.e. $\Psi_1$ and
$\Psi_2$ satisfy the Siegel gauge condition:
\beq
\vev{\Psi_1,Q_B\Psi_2} = h\vev{\phi_1|\phi_2}\vev{G_1|G_2},
\eeq
where $h$ is the dimension of $\Psi_1$.

Next we explain how to calculate the cubic terms in the action. We
adopt the method given in \cite{rz}. Let ${\cal K}$ be an operator and
$v(z)$ a function which transforms as $f\circ {\cal K}(z)
=(f'(z))^h{\cal K}(f(z))+g(f(z),z)$ and $f\circ 
v(z)=(f'(z))^{1-h}v(f(z))$ under the conformal
transformation $f$. One can think of ${\cal K}$ as $T$, $j$(ghost
number current), $b$, $c$, $J^a$(WZW current) and so on. $g(f(z),z)$
represents anomaly in the case of ${\cal K}=T$ or $j$. We take
$g(f(z),z)$ to be zero if ${\cal K}$ is Grassmann odd, and 
assume
$v(z)$ to be regular at infinity. Then the conservation law for 
${\cal K}$ is (for the notation see \cite{rz})
\bea
0 & = & \vev{\left(\oint_{\cal C}\frac{dz}{2\pi i}v(z){\cal K}(z)
-\oint_{\wt{z}=0}\frac{d\wt{z}}{2\pi i}g(I(\wt{z}),\wt{z})
I\circ v(\wt{z})\right)
f_1\circ\Phi_1(0)f_2\circ\Phi_2(0)f_3\circ\Phi_3(0)}\nn\\
& = & \Biggl\langle\left[\sum_{i=1}^3\left(\oint_{{\cal C}_i}
\frac{dz_i}{2\pi i}f_i\circ v(z_i)f_i\circ {\cal K}(z_i)
-g(f_i(z_i),z_i)f_i\circ v(z_i)\right)
-\oint_{\wt{z}=0}\frac{d\wt{z}}{2\pi i}
g(I(\wt{z}),\wt{z})I\circ v(\wt{z})\right] \nn\\
& & \times f_1\circ\Phi_1(0)f_2\circ\Phi_2(0)f_3\circ\Phi_3(0)\Biggr\rangle.
\eea
The representation of this law by oscillators is
\beq
\vev{\left(\sum_{i=1}^3\sum_n{\cal K}_n^{(i)}v_{-n}^{(i)}-A(v)\right)
\Phi_1,\Phi_2,\Phi_3}=0,
\label{conservation}
\eeq
where
\bea
f_i\circ {\cal K}(z_i) & = & \sum_n {\cal K}_n^{(i)}z_i^{-n-h}, \\
f_i\circ v(z_i) & = & \sum_n v_n^{(i)}z_i^{-n+h-1},
\eea
and
\bea
A(f,v) & = & \oint_{z=0}\frac{dz}{2\pi i}g(f(z),z)f\circ v(z),\\
A(v) & = & \sum_{i=1}^3A(f_i,v)+A(I,v).
\eea
We take $v(z)=z^{-l}(z^2-3)^k$ \cite{rz}.
By using $f_2(z_2)=-f_2(-z_2)$ and $f_3(z_3)=-f_1(-z_1)|_{z_1=z_3}$,
we can derive $v^{(2)}_{l+h+2n}=0,\quad v^{(3)}_{-k+h+n}=
(-1)^{k+l+n+1}v^{(1)}_{-k+h+n}\quad (n\in{\bf Z})$,
and from the explicit form of $v(z)$ the Laurent expansions of 
$f_1\circ v(z_1),f_2\circ v(z_2)$ and $f_3\circ v(z_3)$ start from
$z_1^k,z_2^{-l}$ and $z_3^k$ respectively.
We suppose $A(v)=0$ if $l+h=$ even. This is satisfied in the case of
${\cal K}=T$ or $j$. 
Then we can use the relation
\bea
{\cal K}^{(2)}_{-l-h+1} & = & -(v^{(2)}_{l+h-1})^{-1}[-A(v)\delta_{l+h,odd}
+\sum_{m=0}^\infty v^{(2)}_{l+h-3-2m}{\cal K}^{(2)}_{-l-h+3+2m} \nn\\
& & +\sum_{m=0}^\infty v^{(1)}_{-k+h-1-m}({\cal K}^{(1)}_{k-h+1+m}
+(-1)^{k+l-m}{\cal K}^{(3)}_{k-h+1+m})],
\label{modecons}
\eea
in the 3-point function, where $\delta_{l+h,odd}=1 
( \mbox{for } l+h \mbox{ odd}),=0 
( \mbox{for } l+h \mbox{ even or noninteger})$.
If we take $l\geq(>) 1-h$ and $k>(\geq) h-1$, mode numbers of ${\cal K}$ which
appear in the right hand side of (\ref{modecons}) are greater than 
that of the left hand side. Hence by using (\ref{modecons}) iteratively, 
3-point functions of descendants can be reduced to those of primaries,
which are given in the SU(2) case as follows.
\bea
\vev{\Phi_{j_1}^{m_1},\Phi_{j_2}^{m_2},\Phi_{j_3}^{m_3}} 
& = & \left(\frac{8}{3}\right)^{h_1+h_3}\left(\frac{2}{3}\right)^{h_2}
\vev{\Phi_{j_1}^{m_1}(\sqrt{3})\Phi_{j_2}^{m_2}(0)
 \Phi_{j_3}^{m_3}(-\sqrt{3})} \nn\\
& = & K^{-h_{j_1}-h_{j_2}-h_{j_3}}
 C\left[\begin{array}{ccc}
  m_1 & m_2 & m_3 \\
  j_1 & j_2 & j_3 \\
 \end{array}\right],
\eea
where $K=\frac{3\sqrt{3}}{4}$.

$\vev{\Psi_1,\Psi_2,\Psi_3}$ has the cyclic symmetry:
$\vev{\Psi_1,\Psi_2,\Psi_3}=(-1)^{\gamma^3(\gamma^1+\gamma^2)}
\vev{\Psi_3,\Psi_1,\Psi_2}$, where $(-1)^{\gamma^i}$ is the 
Grassmann parity of $\Psi_i$. In addition
to this, there is the following relation:
if $\phi_i$ have Grassmann parity $(-1)^{g_i}$ and are level $N_i$ 
descendants of the primary operators
$\Phi_i$ with Grasmann parity $(-1)^{G_i}$, and satisfy the relation
$\vev{\Phi_1,\Phi_2,\Phi_3}=(-1)^{N_{123}+G_1G_2+G_2G_3+G_3G_1}
\vev{\Phi_3,\Phi_2,\Phi_1}$, then
\beq
\vev{\phi_1,\phi_2,\phi_3}=(-1)^{N_{123}+N_1+N_2+N_3
 +g_1g_2+g_2g_3+g_3g_1}\vev{\phi_3,\phi_2,\phi_1}.
\label{exsym}
\eeq
We can regard this relation as the result of worldsheet parity symmetry,
and this can also be proven directly as follows by applying the above 
procedure. We use the induction on the descendant level of $\phi_{1,2,3}$:
\bea
-v^{(2)}_n\vev{\phi_1,{\cal K}_{-n}\phi_2,\phi_3} & = & 
-A(v)\delta_{n,even}\vev{\phi_1,\phi_2,\phi_3} \nn\\
& & +\sum_{m=0}^\infty\Big[ 
 v^{(2)}_{n-2m-2}\vev{\phi_1,{\cal K}_{-n+2m+2}\phi_2,\phi_3} \nn\\
& & +(-1)^{g_{\cal K}g_1}v^{(1)}_{-k+h-1-m}
 \vev{{\cal K}_{k-h+1+m}\phi_1,\phi_2,\phi_3} \nn\\
& & +(-1)^{k-h+1+n+m+g_{\cal K}g_2}v^{(1)}_{-k+h-1-m}
 \vev{\phi_1,\phi_2,{\cal K}_{k-h+1+m}\phi_3} \Big] \nn\\
& = & -(-1)^{N_{123}+N_1+N_2+N_3+g_1g_2+g_2g_3+g_3g_1}
 A(v)\delta_{n,even}\vev{\phi_3,\phi_2,\phi_1} \nn\\
& & +\sum_{m=0}^\infty\Big[ 
 (-1)^{N_{123}+N_1+N_2+N_3+g_1(g_2+g_{\cal K})
 +(g_2+g_{\cal K})g_3+g_3g_1+n-2m-2} \nn\\
& &  \times v^{(2)}_{n-2m-2}
 \vev{\phi_3,{\cal K}_{-n+2m+2}\phi_2,\phi_1} \nn\\
& & +(-1)^{N_{123}+N_1+N_2+N_3
 +(g_1+g_{\cal K})g_2+g_2g_3+g_3(g_1+g_{\cal K})
 +g_{\cal K}g_1-k+h-1-m} \nn\\
& & \times v^{(1)}_{-k+h-1-m}
 \vev{\phi_3,\phi_2,{\cal K}_{k-h+1+m}\phi_1} \nn\\
& & +(-1)^{N_{123}+N_1+N_2+N_3
 +g_1g_2+g_2(g_3+g_{\cal K})+(g_3+g_{\cal K})g_1+n+g_{\cal K}g_2} \nn\\
& & \times v^{(1)}_{-k+h-1-m}
 \vev{{\cal K}_{k-h+1+m}\phi_3,\phi_2,\phi_1} \Big] \nn\\ 
& = & (-1)^{N_{123}+N_1+(N_2+n)+N_3+g_1(g_2+g_{\cal K})
 +(g_2+g_{\cal K})g_3+g_3g_1}
 \Bigg[-A(v)\delta_{n,even}\vev{\phi_3,\phi_2,\phi_1} \nn\\
& & +\sum_{m=0}^\infty\Big[
 v^{(2)}_{n-2m-2}\vev{\phi_3,{\cal K}_{-n+2m+2}\phi_2,\phi_1} \nn\\
& & +(-1)^{g_{\cal K}g_3}v^{(1)}_{-k+h-1-m}
 \vev{{\cal K}_{k-h+1+m}\phi_3,\phi_2,\phi_1} \nn\\ 
& & +(-1)^{g_{\cal K}g_2-k+h-1-n-m}v^{(1)}_{-k+h-1-m}
 \vev{\phi_3,\phi_2,{\cal K}_{k-h+1+m}\phi_1} \Big]\Bigg] \nn\\
& = & -v^{(2)}_n 
 (-1)^{N_{123}+N_1+(N_2+n)+N_3+g_1(g_2+g_{\cal K})
 +(g_2+g_{\cal K})g_3+g_3g_1}
 \vev{\phi_3,{\cal K}_{-n}\phi_2,\phi_1}. 
\eea
Here we used the assumption of the induction and $A(v)=0$ when $g_{\cal
 K}(\mbox{Grasmann parity of } {\cal K})=1$.
In addition, by using the cyclic symmetry,
\bea
\vev{\phi_1,\phi_2,{\cal K}_{-n}\phi_3}
& = & (-1)^{N_{123}+N_1+N_2+(N_3+n)
 +g_1g_2+g_2(g_3+g_{\cal K})+(g_3+g_{\cal K})g_1}
 \vev{{\cal K}_{-n}\phi_3,\phi_2,\phi_1}, \\
\vev{{\cal K}_{-n}\phi_1,\phi_2,\phi_3}
& = & (-1)^{N_{123}+(N_1+n)+N_2+N_3
 +(g_1+g_{\cal K})g_2+g_2g_3+g_3(g_1+g_{\cal K})}
 \vev{\phi_3,\phi_2,{\cal K}_{-n}\phi_1}.
\eea
Hence we conclude that (\ref{exsym}) is valid by induction.
We can use (\ref{exsym}) and the cyclic symmetry to reduce the number 
of terms in the action necessary for our calculation.
In the SU(2) case, $N_{123}=j_1+j_2+j_3$ for $\Phi_i=\Phi_{j_i}^{m_i}$, 
as can be seen from (\ref{ccyc}).

As an example, let us calculate 
$\vev{J_{-1}^a\Phi_{j_1}^{m_1},J_{-1}^b\Phi_{j_2}^{m_2},\Phi_{j_3}^{m_3}}$
by using the eq.(4.18) of \cite{rz} for WZW current $J^a$: 
$J^{(2)a}_{-1}=-(\frac{5}{27}J^{(2)a}_1+\cdots)
 +(-\frac{2}{3\sqrt{3}}J^{(1)a}_0+\frac{16}{27}J^{(1)a}_1+\cdots)
 +(\frac{2}{3\sqrt{3}}J^{(3)a}_0+\frac{16}{27}J^{(3)a}_1+\cdots)$.
\bea
& & \vev{J_{-1}^a\Phi_{j_1}^{m_1},J_{-1}^b\Phi_{j_2}^{m_2},\Phi_{j_3}^{m_3}}
\nn\\
& = & -\frac{2}{3\sqrt{3}}
 \vev{J_0^bJ_{-1}^a\Phi_{j_1}^{m_1},\Phi_{j_2}^{m_2},\Phi_{j_3}^{m_3}}\nn\\
& & +\frac{16}{27}
 \vev{J_1^bJ_{-1}^a\Phi_{j_1}^{m_1},\Phi_{j_2}^{m_2},\Phi_{j_3}^{m_3}}\nn\\
& & +\frac{2}{3\sqrt{3}}
 \vev{J_{-1}^a\Phi_{j_1}^{m_1},\Phi_{j_2}^{m_2},J_0^b\Phi_{j_3}^{m_3}}\nn\\
& = & -\frac{2}{3\sqrt{3}}if^{bac}
 \vev{J_{-1}^c\Phi_{j_1}^{m_1},\Phi_{j_2}^{m_2},\Phi_{j_3}^{m_3}}
 -\frac{2}{3\sqrt{3}}
 \vev{J_{-1}^a\Phi_{j_1}^{m'_1},\Phi_{j_2}^{m_2},\Phi_{j_3}^{m_3}}
 (T^b_{(j_1)})_{m'_1}^{\;\;m_1} \nn\\
& & +\frac{16}{27}if^{bac}
 \vev{J_0^c\Phi_{j_1}^{m_1},\Phi_{j_2}^{m_2},\Phi_{j_3}^{m_3}}
 +\frac{16}{27}k\delta^{ab}
 \vev{\Phi_{j_1}^{m_1},\Phi_{j_2}^{m_2},\Phi_{j_3}^{m_3}}\nn\\
& & +\frac{2}{3\sqrt{3}}
 \vev{J_{-1}^a\Phi_{j_1}^{m_1},\Phi_{j_2}^{m_2},\Phi_{j_3}^{m'_3}}
 (T^b_{(j_3)})_{m'_3}^{\;\;m_3} \nn\\
& = & -\frac{2}{3\sqrt{3}}if^{bac}\left(\frac{2}{3\sqrt{3}}\right)
 \vev{\Phi_{j_1}^{m_1},J_0^c\Phi_{j_2}^{m_2},\Phi_{j_3}^{m_3}}
 -\frac{2}{3\sqrt{3}}if^{bac}\left(-\frac{2}{3\sqrt{3}}\right)
 \vev{\Phi_{j_1}^{m_1},\Phi_{j_2}^{m_2},J_0^c\Phi_{j_3}^{m_3}}\nn\\
& & -\frac{2}{3\sqrt{3}}\left(\frac{2}{3\sqrt{3}}\right)
 \vev{\Phi_{j_1}^{m'_1},J_0^a\Phi_{j_2}^{m_2},\Phi_{j_3}^{m_3}}
 (T^b_{(j_1)})_{m'_1}^{\;\;m_1}
 -\frac{2}{3\sqrt{3}}\left(-\frac{2}{3\sqrt{3}}\right)
 \vev{\Phi_{j_1}^{m'_1},\Phi_{j_2}^{m_2},J_0^a\Phi_{j_3}^{m'_3}} 
 (T^b_{(j_1)})_{m'_1}^{\;\;m_1} \nn\\
& & +\frac{16}{27}if^{bac}
 \vev{\Phi_{j_1}^{m'_1},\Phi_{j_2}^{m_2},\Phi_{j_3}^{m_3}}
 (T^c_{(j_1)})_{m'_1}^{\;\;m_1}
 +\frac{16}{27}k\delta^{ab}
 \vev{\Phi_{j_1}^{m_1},\Phi_{j_2}^{m_2},\Phi_{j_3}^{m_3}}\nn\\
& & +\frac{2}{3\sqrt{3}}\left(\frac{2}{3\sqrt{3}}\right)
 \vev{\Phi_{j_1}^{m_1},J_0^a\Phi_{j_2}^{m_2},\Phi_{j_3}^{m'_3}}
 (T^b_{(j_3)})_{m'_3}^{\;\;m_3}
 +\frac{2}{3\sqrt{3}}\left(-\frac{2}{3\sqrt{3}}\right)
 \vev{\Phi_{j_1}^{m_1},\Phi_{j_2}^{m_2},J_0^a\Phi_{j_3}^{m'_3}} 
 (T^b_{(j_3)})_{m'_3}^{\;\;m_3} \nn\\
& = & \frac{4}{27}\vev{\Phi_{j_1},\Phi_{j_2},\Phi_{j_3}}
 [ T_3^aT_1^b-T_2^aT_1^b+T_2^aT_3^b-T_3^aT_3^b
  +if^{abc}(T_2^c-T_3^c-4T_1^c)+4k\delta^{ab} ].
\eea
In the last line we suppressed the indices $m$ and denote
 $T^a_{(j_i)}$ by $T^a_{i}$. 
At first sight this result does not satisfy the relation (\ref{exsym}). 
However we can rewrite it to an explicitly symmetric form by using the Ward
identity
 $\vev{\Phi_{j_1},\Phi_{j_2},\Phi_{j_3}}(T_1^a+T_2^a+T_3^a)=0$:
\beq
\vev{J_{-1}^a\Phi_{j_1}^{m_1},J_{-1}^b\Phi_{j_2}^{m_2},\Phi_{j_3}^{m_3}}
= -\frac{4}{27}\vev{\Phi_{j_1},\Phi_{j_2},\Phi_{j_3}}
 [ 2T_1^aT_1^b+2T_2^bT_2^a+4T_1^bT_2^a+T_2^bT_1^a
  +2if^{abc}(T_1^c-T_2^c)-4k\delta^{ab} ].
\label{3pointex}\\
\eeq

We collect the results of the calculation of cubic terms in appendix 
\ref{appb}
and the terms in the action necessary for our calculation are in 
appendix \ref{appc}.

\section{Lump Solutions \label{sec3}}

In this section we construct lump solutions and compare their energies
with the expected values. We denote by [$N$,$M$] the truncation that
we consider only fields with 
the level (i.e. conformal dimension$+1$) $<N$ and retain only 
the terms with the sum of the levels $<M$. 

The D2-brane corresponding to the boundary state $\ket{J}$ can be
regarded as the bound state of $(2J+1)$ D0-branes \cite{bds,ars2}, 
and D0-branes
are conjectured to be able to vanish into the vacuum by tachyon
condensation. Therefore one can imagine the process that some of the
D0-branes forming the D2-brane vanish and the other D0-branes remain.
Hence it is natural to conjecture that there exist lump solutions on
D2-branes which represent $n$ D0-branes $(n<2J+1)$. 
The energy $V({\cal T})$ of a classical configuration ${\cal T}$ is
minus the action:
\bea
V({\cal T})& = & -S({\cal T})=\frac{1}{g_o^2}(-g_o^2S({\cal T}))
 =2\pi^2 M_J(-g_o^2S({\cal T})) \nn\\
 & \equiv & M_J f({\cal T}).
\eea
For the solution ${\cal T}_{\rm vac}$ corresponding to the vacuum, it
is conjectured that $f({\cal T}_{\rm vac})=-1$. It can be expected
that there exist the solutions ${\cal T}_n$ representing $n$ D0-branes
(in this sense ${\cal T}_{\rm vac}={\cal T}_0$) and the difference
between $V({\cal T}_n)$ and $V({\cal T}_{\rm vac})$ is equal to
$nM_0$:
\bea
V({\cal T}_n) & = & V({\cal T}_{\rm vac})+nM_0
 \nn\\ 
 & = & M_J\left(f({\cal T}_{\rm vac})+n\cdot\frac{M_0}{M_J}\right)
 \nn\\ 
 & = & M_J\left(f({\cal T}_{\rm vac})+
 n\cdot\frac{\sin{\frac{\pi}{k+2}}}{\sin{\frac{(2 J +1)\pi}{k+2}}}\right).
\eea
Hence,
\beq
f({\cal T}_n)=f({\cal T}_{\rm vac})+
 n\cdot\frac{\sin{\frac{\pi}{k+2}}}{\sin{\frac{(2 J +1)\pi}{k+2}}}
 = -1 +n\cdot\frac{\sin{\frac{\pi}{k+2}}}{\sin{\frac{(2 J +1)\pi}{k+2}}}.
\eeq
We denote $f$ and ${\cal T}_n$ calculated in level [$N$,$M$]
truncation by $f^{[N,M]}$ and ${\cal T}_n^{[N,M]}$, and define
$R^{[N,M]}_n$ as follows.
\beq
R^{[N,M]}_n\equiv \frac{f^{[N,M]}({\cal T}_n^{[N,M]})}{f({\cal T}_n)}.
\eeq
$R^{[N,M]}_n$ is a function of $k$. We will calculate $R^{[N,M]}_n$ at 
various values of $k$ and see to what extent it is close to 1.

Note that the configuration given by 
SU(2) transformation (generated by $J_0^a$) of a solution is also a 
solution. For lump
solutions, these degrees of freedom correspond to the positions of
D0-branes.

We will consider the cases with small $J$.
If $J=0$ (i.e. D0-brane) there are no nontrivial primary operators and 
it is impossible to construct lump solution.

Then we will consider the following two cases.
\begin{itemize}
\item $J=1/2, k\geq 2$
\item $J=1, k\geq 4$
\end{itemize}
In the first case, primary operators on the D2-brane are $\Phi_0^0$
and $\Phi_1^m$. (If $k=1$ only $\Phi_0^0$ appears.)
Their dimensions are $h_0=0, h_1=\frac{2}{k+2}$. This case corresponds 
to the bound state of 2 D0-branes and we expect the existence of the
solution ${\cal T}_1$.

In the second case, primary operators are $\Phi_0^0, \Phi_1^m$ and
$\Phi_2^m$. (If $k\leq 3$ some of them do not appear and this case
reduces to the previous one.) Their dimensions are $h_0, h_1, 
h_2=\frac{6}{k+2}$. This case corresponds 
to the bound state of 3 D0-branes and we expect the existence of the
solutions ${\cal T}_1$ and ${\cal T}_2$.

We calculate the action in level [1,3], [2,6] and [5/2,5]. 
The actual calculation is done by symbolic manipulation program
Mathematica. 
\subsection{level [1,3], $k\rightarrow\infty$, $J=$fixed}
In this case $\Psi=\ket{j,m} t_m^j$. The action is 
\beq
\frac{f^{[1,3]}}{2\pi^2}
=\Half\frac{(-1)^m}{\sqrt{(2j+1)[2j+1]}}(h_j-1)t_{-m}^jt_m^j
+\frac{1}{3}K^{3-h_{j_1}-h_{j_2}-h_{j_3}}
 C\left[\begin{array}{ccc}
  m_1 & m_2 & m_3 \\
  j_1 & j_2 & j_3 \\
 \end{array}\right]t^{j_1}_{m_1}t^{j_2}_{m_2}t^{j_3}_{m_3}.
\eeq
First, let us consider the case where $k\rightarrow\infty$ and $J=$
fixed. In this case it has been shown \cite{ars1} that the structure of 
fuzzy spheres appears as follows.
The spectrum of primary operator is $\Phi^m_j,\quad j=0,1,\cdots 2J$
and the number of degrees of freedom of $t_m^j$ is $1+3+\cdots+(2\cdot 
2J+1)=(2J+1)\times(2J+1)$. These form a $(2J+1)\times(2J+1)$ matrix
$t$:
\beq
(t)_{m_1m_2}=\sum_{j,m}\sqrt{\frac{2J+1}{2j+1}}(-1)^jt_m^j(T_{j,m})_{m_1m_2},
\label{matrix}
\eeq
and \cite{h}
\beq
(T_{j,m})_{m_1m_2}=(-1)^{J-m_2}\sqrt{2j+1}\vev{J,m_2,J,-m_1|J,J,j,-m}.
\eeq
Then the terms in the action are realized as follows.
\bea
\frac{1}{2J+1}{\rm tr}(ttt) & = & \sum_{j_1,j_2,j_3,m_1,m_2,m_3}
 C\left[\begin{array}{ccc}
  m_1 & m_2 & m_3 \\
  j_1 & j_2 & j_3 \\
 \end{array}\right]_{k\rightarrow\infty}
 t^{j_1}_{m_1}t^{j_2}_{m_2}t^{j_3}_{m_3},\nn\\
\frac{1}{2J+1}{\rm tr}(tt) & = &  \sum_{j_1,j_2,m_1,m_2}
 C\left[\begin{array}{ccc}
  m_1 & 0 & m_2 \\
  j_1 & 0 & j_2 \\
 \end{array}\right]_{k\rightarrow\infty}
 t^{j_1}_{m_1}t^{j_2}_{m_2}.
\eea
Hence,
\beq
\frac{f^{[1,3]}}{2\pi^2}=\frac{1}{2J+1}{\rm tr}\left(-\Half tt
+\frac{1}{3}K^3ttt\right).
\eeq
Note that the form appearing in the trace is the same as the tachyon
potential in level (0,0).
Solutions of the equation of motion $-t+K^3t^2=0$ for this action can
be constructed by introducing projection operators as in \cite{ncsoliton}:
\beq
t=\lambda P,\quad P^2=P,\quad {\rm tr}P=(2J+1)-n,
\eeq
and $\lambda$ is the nontrivial solution of
$-\lambda+K^3\lambda^2=0$ i.e. $\lambda=K^{-3}$. 
$P^2=P$ and ${\rm tr}P=2J+1-n$ means that if we diagonalize $P$ then
$2J+1-n$ eigenvalues are $1$ and $n$ eigenvalues are $0$ and therefore
we can interpret this solution as ${\cal T}_n^{[1,3]}$.
The energy of this solution is
\bea
\frac{f^{[1,3]}({\cal T}_n^{[1,3]})}{2\pi^2} & = & 
\frac{1}{2J+1}\left(-\Half\lambda^2+\frac{1}{3}K^3\lambda^3\right)
 {\rm tr}P \nn\\
 & = & -\frac{1}{2J+1}\cdot\frac{1}{6}K^{-6}(2J+1-n).
\eea
In this case $f({\cal T}_n)=-\frac{1}{2J+1}(2J+1-n)$,
and $R^{[1,3]}_n=\frac{\pi^2}{3K^6}\simeq 0.684616$.

In the following we look for the sequences of solutions 
$R^{[1,3]}_n$ of which at $k\rightarrow\infty$ is 0.684616.

\subsection{level [1,3], $J=1/2$}
\begin{figure}[htdp]
\begin{center}
\leavevmode
\epsfxsize=70mm
\epsfbox{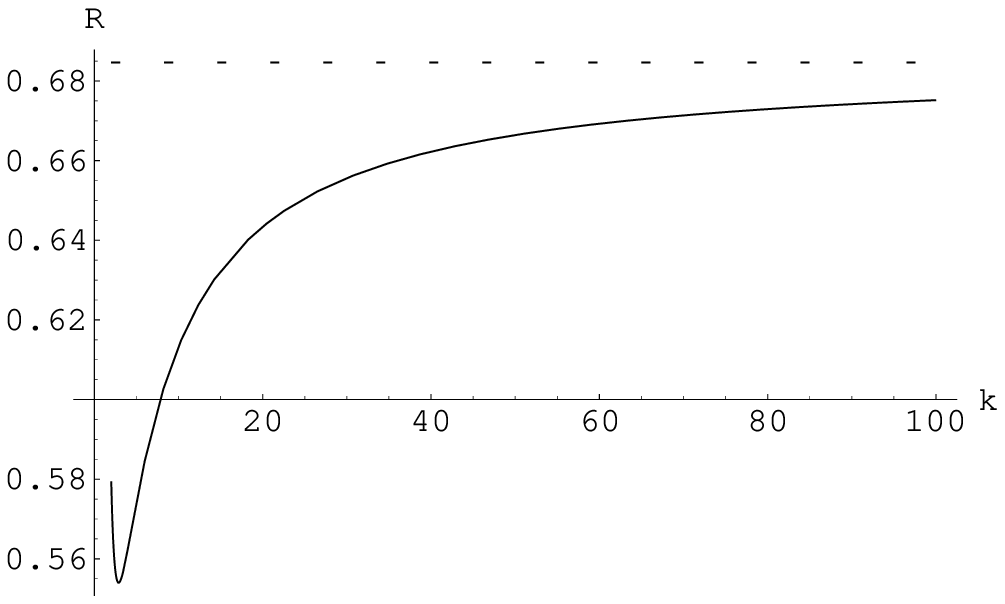}
\caption{$R^{[1,3]}_1$ ($J=\Half$, $k\geq 2$)}  
\label{fig1r}
\leavevmode
\epsfxsize=70mm
\epsfbox{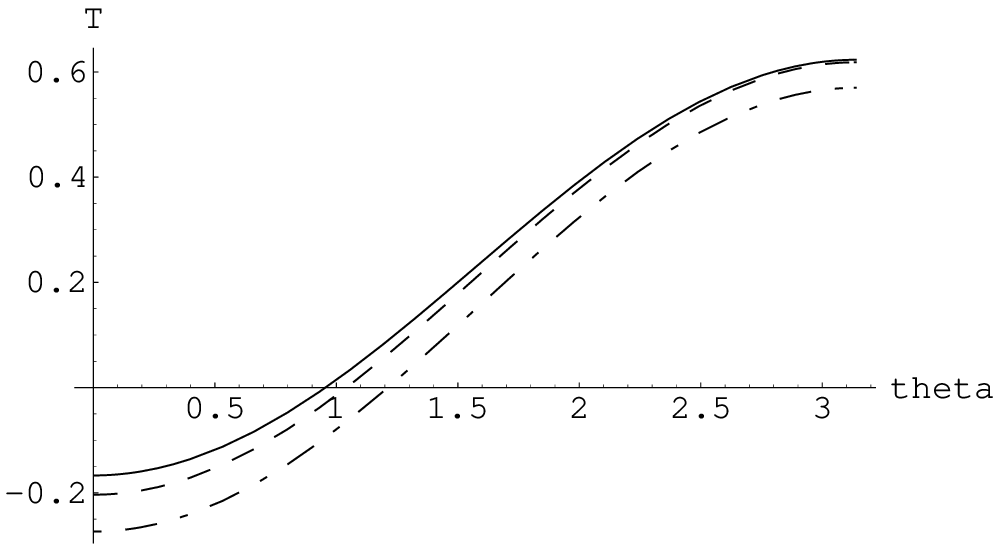}
\caption{
$T(\theta)$ (level [1,3], $J=\Half$) \quad
$k=100000000$ (solid line),
$k=10$ (dashed line),
$k=2$ (dot-dashed line)
}
\label{fig1p}
\end{center}
\end{figure}

In this case we can easily calculate the
action without dropping $J^3_0\neq 0$ components:
\bea
\frac{f^{[1,3]}}{2\pi^2} & = & -\Half (t_0^0)^2
-\frac{k}{k+2}\frac{1}{\sqrt{3[3]}}|t_m^1|^2
+\frac{1}{3}K^3(t_0^0)^3
+\frac{1}{\sqrt{3[3]}}K^{3-4/(k+2)}|t_m^1|^2t_0^0.
\eea
We can easily solve the equations of motion and we find two
solutions. One solution corresponds to the closed string vacuum, in
which the components corresponding to nontrivial primary operators and
their descendants vanishes. The other solution is
\bea
t_0^0 & = & \frac{k}{2(k+2)}K^{4/(k+2)-3},\nn\\
|t_m^1|^2 & = & \sqrt{3 [ 3]}\frac{k}{2(k+2)}K^{8/(k+2)-6}
 \left(1-\frac{k}{2(k+2)}K^{4/(k+2)}\right),
\eea
which is correspond to ${\cal T}_1^{[1,3]}$.
\beq
\frac{f^{[1,3]}({\cal T}_1^{[1,3]})}{2\pi^2}
=\frac{k^2}{4(k+2)^2}K^{8/(k+2)-6}
 \left(-\Half+\frac{k}{6(k+2)}K^{4/(k+2)}\right).
\eeq
$R^{[1,3]}_1$ is shown in fig.\ref{fig1r}. 
The value for $k\rightarrow\infty$ is 0.684616, which agrees with the
value expected above. The minimum is 0.554151 at $k=3$.

By SU(2) rotation, we can put $t_{\pm 1}^1=0$. Then
$T(\theta)\equiv\sqrt{\frac{4\pi}{(2j+1)^{1/2}[2j+1]^{1/2}}}
(-1)^jt_m^jY_j^m(\theta)$ is
not dependent on $\phi$, and its profile is shown in fig.\ref{fig1p}. 
Here we choose $t_0^1$ positive.
(The factor $\sqrt{\frac{4\pi}{(2j+1)^{1/2}[2j+1]^{1/2}}}(-1)^j$ is 
present because $\Phi^m_j$ is corresponding to 
$\sqrt{\frac{4\pi}{(2j+1)^{1/2}[2j+1]^{1/2}}}(-1)^jY_j^m$,
which is obtained by comparing 
$\int d\Omega (-1)^{m'}Y^{-m'}_{j'}Y^m_j=\delta_{jj'}\delta_{mm'}$
with $\vev{(-1)^{m'}\Phi^{-m'}_{j'}\Phi^m_j}=\delta_{jj'}\delta_{mm'}
\sqrt{\frac{(2j+1)^{1/2}[2j+1]^{1/2}}{4\pi}}$, and referring to 
(\ref{matrix}).)
It seems that this solution is not well localized. But this is not a
problem since the radius of $S^2$ on which the D2-brane is wrapped is
$\sim\sqrt{k}\sin{\frac{2\pi J}{k}}$ and this has the maximum $2\sqrt{J}$
at $k=8J$. This is sufficiently small for small $J$. Therefore the
worldvolume of the D2-brane itself is very small. At any rate we cannot
construct well localized configuration by using only
$Y_0^0=\frac{1}{4\pi},Y_1^0=\frac{3}{4\pi}\cos\theta$ and 
 $Y_1^{\pm 1}=\mp\frac{3}{8\pi}\sin\theta e^{\pm i\theta}$.

If we choose the sign of $t_0^1$ negative, the profile is given 
by $T(\pi-\theta)$. These two configurations seem to represent D0-branes 
at $\theta=0$ and $\pi$.

\subsection{level [1,3], $J=1$}
In this case we put $J^3_0=0$ components zero. Then
\bea
\frac{f^{[1,3]}}{2\pi^2} & = & 
-\Half (t^0)^2
+\Half \frac{1}{\sqrt{3 [3]}} (-1+\frac{2}{2+k}) (t^1)^2
+\Half \frac{1}{\sqrt{5 [5]}} (-1+\frac{6}{2+k}) (t^2)^2 \nn\\
& & +\frac{1}{3} K^3 (t^0)^3 
+ \frac{1}{\sqrt{3 [3]}}K^{3-\frac{4}{k+2}} (t^0) (t^1)^2 
+\frac{1}{\sqrt{5 [5]}}K^{3-\frac{12}{2+k}} (t^0) (t^2)^2 \nn\\
& & +\sqrt{\frac{2}{3}} \frac{1}{\sqrt{5 [5]}} c_{11}^2 
 K^{3-\frac{10}{k+2}} (t^1)^2 (t^2)
-\frac{1}{3} \sqrt{\frac{2}{7}} \frac{1}{\sqrt{5 [5]}} c_{22}^2 
 K^{3-\frac{18}{k+2}} (t^2)^3.
\eea
If $k=4$, $t^2$ must be put zero since the dimension of $\Phi_2^m$ is $1$.
In the case with $k\neq 4$, 
we can reduce the equations of motion to quadratic equations, and for 
$k\geq6$ we can find essentially two solutions which have 
$R^{[1,3]}_1\simeq 0.684616$ and
$R^{[1,3]}_2\simeq 0.684616$ at $\infty$ respectively.
The case with $k=4$ and $5$ is somewhat special. we find only one 
nontrivial
solution, which seems to be correspond to ${\cal T}_1^{[1,3]}$
 ($R^{[1,3]}_1=0.475766$ for $k=4$ and $R^{[1,3]}_1=0.641119$ for $k=5$). 
We do not find any solution corresponding to ${\cal T}_2^{[1,3]}$. 

\begin{figure}[htdp]
\begin{center}
\leavevmode
\epsfxsize=70mm
\epsfbox{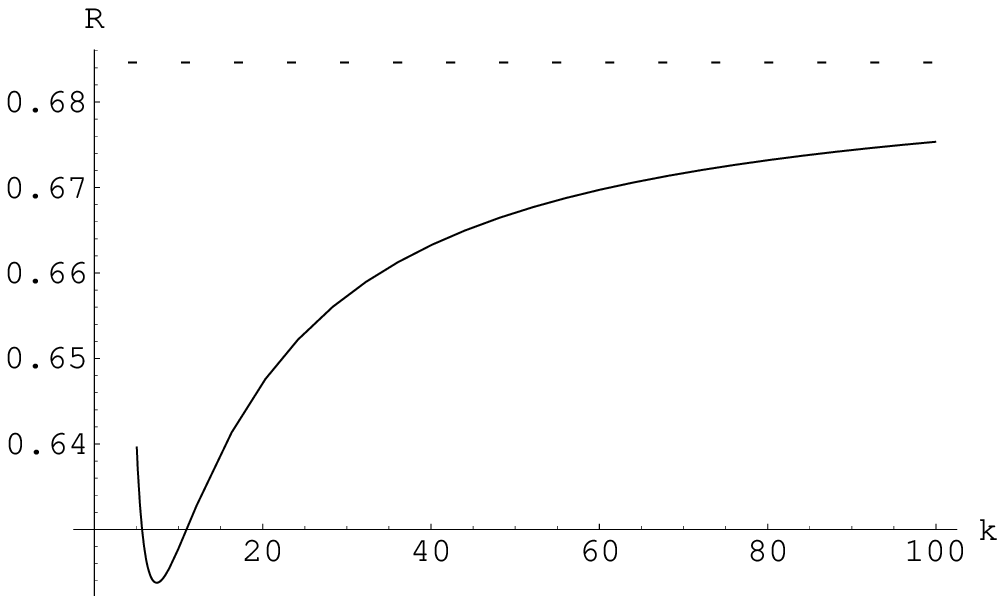}
\caption{$R^{[1,3]}_1$ ($J=1$, $k\geq 4$)}
\label{fig2r}
\leavevmode
\epsfxsize=70mm
\epsfbox{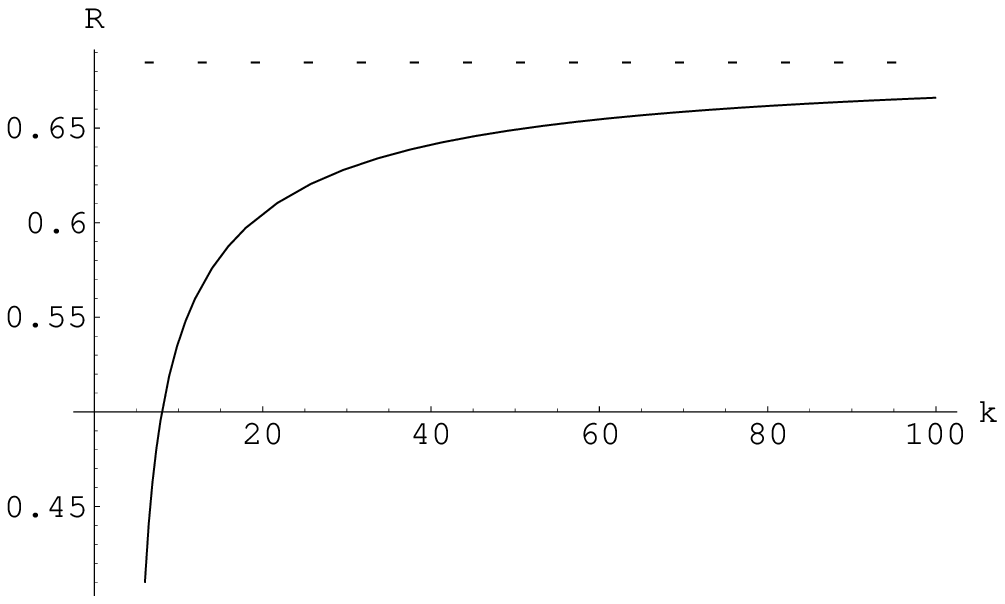}
\caption{$R^{[1,3]}_2$ ($J=1$, $k\geq 6$)}
\label{fig3r}
\leavevmode
\epsfxsize=70mm
\epsfbox{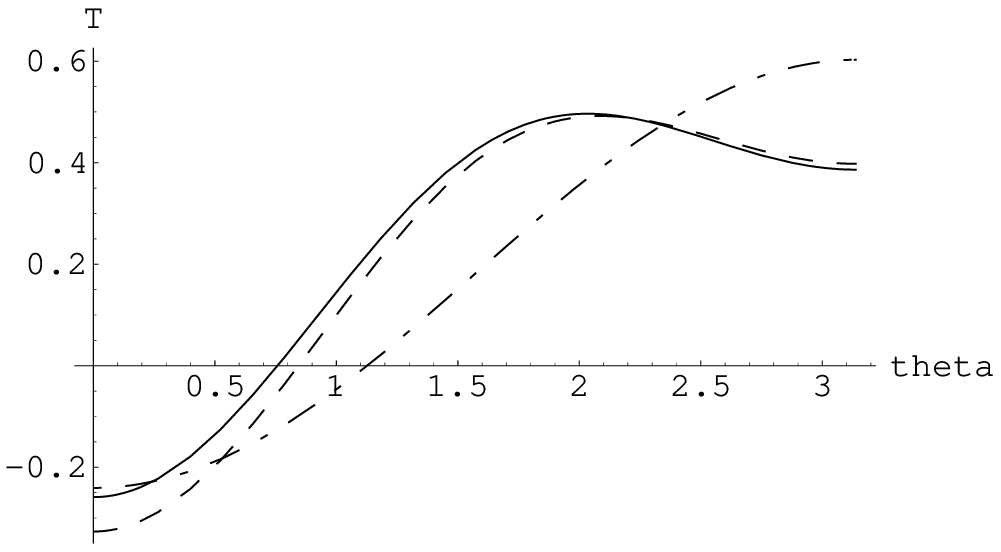}
\caption{$T(\theta)$ (level [1,3], $J=1$, ${\cal T}_1$) \quad
$k=100000000$ (solid line),
$k=10$ (dashed line),
$k=4$ (dot-dashed line)
}
\label{fig2p}
\leavevmode
\epsfxsize=70mm
\epsfbox{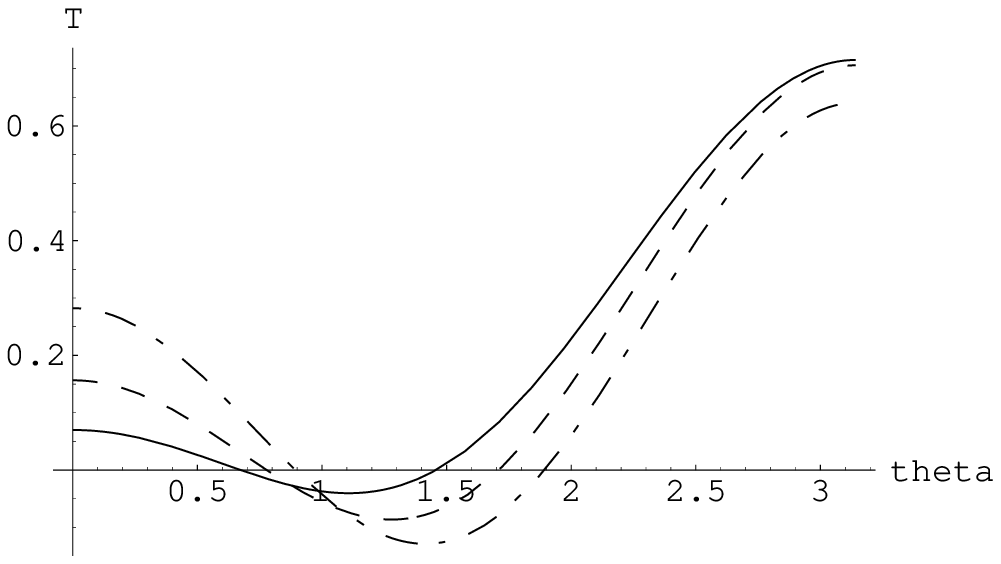}
\caption{$T(\theta)$ (level [1,3], $J=1$, ${\cal T}_2$) \quad
$k=100000000$ (solid line),
$k=10$ (dashed line),
$k=6$ (dot-dashed line)
}
\label{fig3p}
\end{center}
\end{figure}

$R^{[1,3]}_{1,2}$ and the profiles of 
$T(\theta)\equiv\sqrt{\frac{4\pi}{(2j+1)^{1/2}[2j+1]^{1/2}}}
(-1)^jt_m^jY_j^m(\theta)$ are shown in fig.
\ref{fig2r},\ref{fig3r},\ref{fig2p},\ref{fig3p}.
The minimum of $R^{[1,3]}_1$ and $R^{[1,3]}_2$ is 0.623983 at $k=7$ and 
0.409721 at $k=6$ respectively.

\subsection{level [2,6]}
In this case, we can put $\Psi=\ket{j,m}t_m^j + J_{-1}^a\ket{j,m} u_m^{aj}
+c_0b_{-1}\ket{j,m}\wt{t}_m^j $.
The action is too long and it is not illuminating to write it here.
Therefore we describe only its salient features.

$\wt{t}_m^j$ can be taken to be zero in the Siegel gauge. However their 
equations of motion may give nontrivial constraints \cite{hsms}. 
The linear term in $\wt{t}_m^j$ in the action is 
\beq
\frac{(-1)^m}{\sqrt{(2j+1) [ 2j+1]}}
 \wt{t}^{j}_{-m}(T^a_j)_m^{\;\; m'}u_{m'}^{aj}.
\eeq
This term vanishes in our case since 
$(T^a_j)_0^{\;\; m'}u_{m'}^{aj}\sim u^{jj}=0$ 
as can be seen from the results below. 
Therefore we can put $\wt{t}_0^j=0$ to satisfy their equations of motion
and can adopt Siegel gauge condition for looking for lump solutions.

In this case we cannot expect any improvement at $k\rightarrow\infty$
comparing with the results of level [1,3].
The reason is as follows. 
The linear term in $u_m^{aj}$ is
\beq
\frac{2}{3\sqrt{3}}K^{3-h_{j_1}-h_{j_2}-h_{j_3}}
 C\left[\begin{array}{ccc}
  m_1 & m_2 & m_3 \\
  j_1 & j_2 & j_3 \\
 \end{array}\right]
((T^a_{j_3})_{m_3}^{\;\; m'_3}\delta_{m_1}^{\;\; m'_1}-
 (T^a_{j_1})_{m_1}^{\;\; m'_1}\delta_{m_3}^{\;\; m'_3})
 t^{j_1}_{m'_1}u_{m_2}^{aj_2}t^{j_3}_{m'_3},
\eeq
and at $k\rightarrow\infty$ the most dominant quadratic term in
$u_m^{aj}$ is given from the last term of eq.(\ref{3pointex}):
\beq
\frac{16}{27}k\cdot K^{3-h_{j_1}-h_{j_2}-h_{j_3}}
 C\left[\begin{array}{ccc}
  m_1 & m_2 & m_3 \\
  j_1 & j_2 & j_3 \\
 \end{array}\right]
 u^{aj_1}_{m_1}u_{m_2}^{aj_2}t^{j_3}_{m_3}.
\eeq
(There is also cubic terms $u_m^{aj}$ which have coefficients
proportional to $k$. This kind of term is generated by the second term
of commutation relation (\ref{commutation})).
To let the contribution of this term be finite, we rescale $u_m^{aj}$
to $\frac{1}{\sqrt{k}}u_m^{aj}$. Then the contribution of the linear
term in $u_m^{aj}$ vanishes. Hence we can get the solution by putting
$u_m^{aj}$ zero, and its energy is the same as that of level [0,0].

We give the values obtained in level [1,3] analysis to $t_m^j$ as the
initial values and perform the numerical analysis. The results are 
summarized in the table 1,2 and 3 of Appendix \ref{appd}. In these tables 
the rescaling of $u_m^{aj}$
explained above are explicitly expressed by $\sqrt{k}$ i.e. $u_m^{aj}$
in these tables are unrescaled ones.

At large $k$ the value of $R$ is not improved at all, as is expected
above. At small $k$ there seems to be small improvement. 
In the case of $k=4$ and $5$, we do not find any solution 
corresponding to 
${\cal T}^{[1,3]}_2$. Here we can find a solution corresponding to
${\cal T}^{[2,6]}_2$ for $k=5$.

\subsection{level [5/2,5]}
We do not write the full expression of the action because it is not
illuminating. 

Similarly to the case of level [1,3], we can estimate the value of $R$ 
at $k\rightarrow\infty$. We rescale the fields as follows so that the
most dominant terms in the action give finite contribution.
\bea
u^{jj'} & \rightarrow & \frac{1}{\sqrt{k}} u^{jj'}, \nn\\
v^{jj'} & \rightarrow & \frac{1}{\sqrt{k}} v^{jj'}, \nn\\
v^{jj'j''} & \rightarrow & \frac{1}{k} v^{jj'j''}, \nn\\
\wt{u}^{jj'} & \rightarrow & \frac{1}{\sqrt{k}} \wt{u}^{jj'}.
\eea
Then we collect the terms which are finite in the limit
$k\rightarrow\infty$ and they can be represented by matrices as has
been explained in the previous section. The remaining terms are in
fact obtained by considering only the second terms of the commutation
relation (\ref{commutation}), which is in the same form as that of 
flat CFT oscillators. Therefore if we put all the components 
of the string field proportional to a projection operators $P$, the action 
is equal to $\frac{1}{2J+1}V{\rm tr}P$, where $V$ is the tachyon potential
in level (2,4). $V$ is given in \cite{sz} (in the Siegel gauge) 
and for the vacuum solution it is 
0.948553 times the tension of D-brane. Hence we get the lump solutions with 
$R^{[5/2,5]}=$0.948553. 

We look for the sequences of solutions with $R^{[2,5/2]}_n=0.948553$
at $k\rightarrow\infty$. We take the Siegel gauge to look for 
the solutions. The results are shown in  table 4,5 and 6 of 
Appendix \ref{appd}.
We have checked that the matrices (\ref{matrix}) become projections at
$k\rightarrow\infty$ for these solutions.
The profiles of the solutions are almost the same as those of level [1,3].
Therefore we do not show them.

Now we must check the validity of the Siegel gauge condition \cite{hsms}. 
i.e. the
equations of motion of $\wt{t}^j$, $\wt{u}^j$ and $\wt{u}^{jj'}$.
We need only the terms linear in $\wt{t}^j$, $\wt{u}^j$ and $\wt{u}^{jj'}$.
See appendix \ref{appc}.
Then the equations of motion have linear and quadratic terms in the other 
components.
We see whether the equations of motion for $\wt{t}^j$, $\wt{u}^j$ and 
$\wt{u}^{jj'}$ are satisfied when we put the numerical solutions of 
table. The result is shown in table 7,8 and 9 of Appendix \ref{appd}. 
For the equations of motion of $\wt{t}^j$ and 
$\wt{u}^j$ linear terms and quadratic terms cancel each other and the 
ratios of their sums and the linear terms are around 0.2. For the 
equations of motion of $\wt{u}^{jj'}$ there seems to be no significant 
cancellation, but the numerical values themselves are in the same order as 
those of $\wt{t}^j$ and $\wt{u}^j$, or smaller than them. Indeed there are 
no linear terms in $\wt{u}^{jj'}$ at $k\rightarrow\infty$ as can be seen 
by the argument similar to that in the previous section . 
Therefore each of linear and quadratic terms
is small in this limit and they do not have to cancel each other.
It seems that these results support the validity of the Siegel gauge 
condition.

\subsection{$k\rightarrow\infty$, $J=$ fixed}
We saw in the previous discussion that in the case of $J=$fixed 
and $k\rightarrow\infty$, the components of the string field form 
$(2J+1)\times(2J+1)$ matrices and the solutions can be constructed 
by introducing projection operators. The accuracy of the mass of these 
solution is the same as that of the closed string vacuum. This is true 
for arbitrary levels. Indeed, the commutation relation is in the same form 
as that of the flat CFT oscillators in the limit $k\rightarrow\infty$, 
and $c'\rightarrow 23$. (The size of representation matrices of SU(2) 
do not get large as $k\rightarrow\infty$ since $J$ is fixed.) 
Therefore the computation of the coefficients of 
each terms in the action is completely the same as that of the tachyon 
potential on the flat CFT. The OPE of primary fields yields the structure 
of matrix product. As a result the action is equal to the tachyon 
potential with the components of string field replaced by 
$(2J+1)\times(2J+1)$ matrices. This situation is the same as in the case 
where constant B-field is present on the flat space \cite{ncsoliton,w2}.

\section{Summary and discussions \label{sec4}}
We have constructed lump solutions of D-branes on SU(2) group
manifold by level truncation approximation. Their energy
agree with the expected values with good accuracy. 
It is expected that the accuracy is improved if we include 
higher level terms.
We have considered the cases $J=\Half,1$. It is desirable
to consider the cases with larger $J$, though the calculation is more
cumbersome because the number of primary operators increases.
We have seen that in the limit $k\rightarrow\infty,J=$ fixed, the
string field can be expressed by matrices, and by taking them to be
proportional to a projection matrix the action is written as the
tachyon potential times the trace of the projection matrix. 
This case can also be treated by effective action \cite{hnt}, but our 
method can be used for investigating the case with small $k$.

The results show the tendency that the accuracy is improved as $k$
increases. One of the reason for this fact is that the dimension of the
primary operators decrease and more operators must be
included for fixed truncation level. 
Though for small $k$ we do not always obtain very high accuracy as that of 
$k\rightarrow\infty$, there seems to be large improvement as we include 
more terms with higher level. In the case of $k=4, J=1$ we cannot find
the solution corresponding to ${\cal T}_2$. It is expected that we can
find the solution if we add higher level terms, as in the case of $k=5$.

At present it is not clear in what sense the level truncation is 
an approximation. Our results seem to show it is a meaningful 
approximation in our case.

It is interesting to consider lump solutions of D-branes on SU(2)
group manifold by BSFT. Since the background is curved, it seems to be
difficult to obtain exact results as in the case of flat background.
One of the difficulty is that we must consider too many operators for
satisfying the equations of motion in
contrast to the case of flat background, where we can consider
only the tachyon field up to quadratic order in coordinates.
For the large $k$ case see \cite{cs}.

Recently the solutions corresponding to D-branes on arbitrary 
CFT background in vacuum CSFT are constructed in \cite{rsz}.
If the relation between vacuum CSFT and ordinary CSFT
becomes clear, we can relate them to our solutions and they must agree.

We have considered the tachyon condensation. But 
our method can also be applicable to the gluon condensation, by which 
the opposite phenomena to ours happen i.e. many D0-branes form D2-branes
\cite{ars2}.
It is interesting to investigate it by our method.

\vs{.5cm}
\noindent
{\large\bf Acknowledgments}\\[.2cm]
I would like to thank H.\ Hata, S.\ Shinohara and S.\ Teraguchi
for helpful discussions.

\renewcommand{\theequation}{\Alph{section}.\arabic{equation}}
\appendix
\addcontentsline{toc}{section}{Appendices}
\section{Quantum 6j-symbol \label{appa}}
\setcounter{equation}{0}

The symbol $c_{j_1,j_2}^{j_3}(k,J)$ used in eq.(\ref{ope}) is defined
as follows (e.g. \cite{a-ggs}).
\bea
c_{j_1,j_2}^{j_3}(k,J) & = & 
\left\{\begin{array}{ccc}
  j_1 & j_2 & j_3 \\
  J & J & J \\
   \end{array}\right\} \nn\\
& = & (-1)^{-j_1-j_2-2J}\sqrt{[2j_3+1][2J+1]} \nn\\
& & \times [j_1]! \ [j_2]! \ [j_3]! \nn\\
& & \times\sqrt{\frac{[2J-j_1]! \ [2J-j_2]! \ [2J-j_3]!}
 {[2J+j_1+1]! \ [2J+j_2+1]! \ [2J+j_3+1]!}} \nn\\
& & \times\sqrt{\frac{[j_1+j_2-j_3]! \ [j_2+j_3-j_1]! \ [j_3+j_1-j_2]!}
 {[j_1+j_2+j_3+1]!}} \nn\\
& & \times\sum_{z\geq 0}(-1)^z [z+1]! \Bigl\{ \ [z-j_1-j_2-j_3]! \nn\\
& & \times [z-j_1-2J]! \ [z-j_2-2J]! \ [z-j_3-2J]! \nn\\
& & \times [j_1+j_2+2J-z]! \ [j_2+j_3+2J-z]! \ [j_3+j_1+2J-z]! \Bigr\}^{-1},
\label{sixj}
\eea
where $\{\cdot\}$ is the quantum 6j-symbol and the sum $\sum_{z\geq
0}$ is only over those $z$ such that the arguments of [$\,\cdot\,$] are
nonnegative. [$\,\cdot\,$] is defined as follows.
\beq
[n] = \frac{q^{n/2}-q^{-n/2}}{q^{1/2}-q^{-1/2}},
\eeq
where $q=\exp(\frac{2\pi i}{k+2})$, and
\bea
[n]! & = & [1][2]\cdots [n], \\
{[}0{]}! & = & 1.
\eea
Note that the above expression of $c_{j_1,j_2}^{j_3}(k,J)$ is 
totally symmetric under the
permutations of ($j_1,j_2,j_3$) except the first line.
If $k\rightarrow\infty$, then $[n]\rightarrow n$ and the quantum
6j-symbol becomes ordinary 6j-symbol.

\section{Cubic terms \label{appb}}
\setcounter{equation}{0}

We collect the cubic terms in the action needed for the calculation.
These results are applicable to WZW models for general group.
We adopt the abbreviation like the last line of eq.(\ref{3pointex}).
For example, $\vev{1,J_{-1}^a,1}=\frac{2}{3\sqrt{3}}(T_3^a-T_1^a)=
\frac{2}{3\sqrt{3}}\vev{\Phi^{m1'}_{j1},\Phi^{m2'}_{j2},\Phi^{m3'}_{j3}}
(\delta_{m1'}^{m1}\delta_{m2'}^{m2}(T_{(j_3)}^a)_{m3'}^{\;\; m3}
-(T_{(j_1)}^a)_{m1'}^{\;\; m1}\delta_{m2'}^{m2}\delta_{m3'}^{m3})$

\beq
\vev{1,J_{-1}^a,1}= \frac{2}{3\sqrt{3}}(T_3^a-T_1^a)
\eeq
\beq
\vev{J_{-1}^a,J_{-1}^b,1} = -\frac{4}{27}\Bigr[
2 T_1^aT_1^b-2T_2^bT_2^a+4 T_1^bT_2^a+
T_1^aT_2^b +2if^{abc}(T_1^c-T_3^c) -4k\delta^{ab}\Bigr]
\eeq
\bea
\vev{J_{-1}^a,J_{-1}^b,J_{-1}^c} & = & \frac{8}{81\sqrt{3}}\Bigl[
 8kif^{abc} +4k\delta^{ab}T_1^c-4k\delta^{ca}T_1^b\nn\\
& & +4k\delta^{bc}T_2^a
-4k\delta^{ab}T_2^c+4k\delta^{ca}T_3^b
-4k\delta^{bc}T_3^a \nn\\
& & +4f^{bad}f^{cde}T^e
-4f^{bcd}f^{dae}T_1^e
+4f^{bad}f^{cde}T_2^e \nn\\
& & -4f^{bcd}f^{dae}T_2^e
-4f^{bad}f^{cde}T_3^e
+4f^{bcd}f^{dae}T_3^e \nn\\
& & +4if^{bad}T_1^cT_1^d
-4if^{cad}T_1^bT_1^d
-if^{bcd}T_2^aT_2^d
+if^{bad}T_2^dT_2^c \nn\\
& & -4if^{bcd}T_3^aT_3^d
-if^{cad}T_3^dT_3^b
+if^{bcd}T_1^dT_2^a \nn\\
& & -4if^{bad}T_1^dT_2^c
-if^{bad}T_1^cT_2^d
+if^{cad}T_1^bT_2^d
-if^{bcd}T_1^dT_3^a \nn\\
& & +if^{bcd}T_2^dT_3^a
+4if^{cad}T_1^dT_3^b
+if^{cad}T_2^dT_3^b
+if^{bad}T_1^cT_3^d \nn\\
& & -if^{bad}T_2^cT_3^d
+4if^{bcd}T_2^aT_3^d
+if^{cad}T_1^bT_3^d \nn\\
& & +T_2^aT_2^cT_1^b
-T_3^aT_3^bT_1^c
-T_1^cT_1^bT_2^a
+T_3^aT_3^bT_2^c \nn\\
& & +T_1^cT_1^bT_3^a
-T_1^bT_2^cT_3^a
-T_2^aT_2^cT_3^b
+T_1^cT_2^aT_3^b\Bigr]
\eea
\beq
\vev{1,J_{-2}^a,1}= -\frac{2}{9} T_2^a
\eeq
\bea
\vev{1,J_{-1}^aJ_{-1}^b,1} & = & \frac{1}{27} \Bigl[ 4T_1^bT_1^a
 -4T_1^bT_3^a -4T_1^aT_3^b +4T_3^bT_3^a \nn\\
& & -5if^{abc}T_2^c -5k\delta^{ab}\Bigr]
\eea
\beq
\vev{J_{-1}^a,J_{-2}^b,1} = \frac{4}{81\sqrt{3}}
 \Bigl[ 3T_2^bT_3^a- 3T_2^aT_2^b-16if^{abc}T_1^c+16k\delta^{ab}\Bigr]
\eeq
\bea
\vev{J_{-1}^a,J_{-1}^bJ_{-1}^c,1} & = & \frac{1}{81\sqrt{3}}\Bigl[
32ikf^{abc}
+10k\delta^{bc} T_3^a
-10k\delta^{bc} T_2^a \nn\\
& & +32k\delta^{ac} T_3^b
+32k\delta^{ab} T_3^c
-32k\delta^{ab} T_1^c
-32k\delta^{ac} T_1^b \nn\\
& &
+8f^{bad}f^{cde} T_3^e 
-8f^{bad}f^{cde} T_2^e 
+32f^{bad}f^{cde} T_1^e \nn\\
& & 
-8if^{bad} T_1^cT_3^d 
+8if^{bad} T_1^cT_2^d 
+8if^{bad} T_3^dT_3^c 
-8if^{bad} T_2^dT_3^c \nn\\ 
& & 
-8if^{cad} T_1^bT_3^d 
+8if^{cad} T_1^bT_2^d
-32if^{cad} T_1^dT_1^b \nn\\
& & 
-32if^{bad} T_1^cT_1^d 
+32if^{bad} T_1^dT_3^c 
+10if^{bcd} T_2^dT_3^a 
-10if^{bcd} T_2^aT_2^d \nn\\
& & +8if^{cad} T_3^dT_3^b 
-8if^{cad} T_2^dT_3^b 
+32if^{cad} T_1^dT_3^b \nn\\
& & 
-8 T_1^cT_1^bT_3^a 
+8 T_1^cT_1^bT_2^a
-8 T_1^bT_2^aT_3^c
+8 T_1^bT_3^aT_3^c \nn\\
& & 
-8 T_1^cT_2^aT_3^b
+8 T_1^cT_3^aT_3^b 
-8 T_3^aT_3^cT_3^b 
+8 T_2^aT_3^cT_3^b\Bigr]
\eea
\beq
\vev{J_{-2}^a,J_{-2}^b,1} = \frac{4}{729}
 \Bigl[-192k\delta^{ab}-32if^{bac}T_2^c+32if^{bac}T_3^c-96if^{bac}T_1^c
 +9T_1^aT_2^b\Bigr]
\eeq
\bea
\vev{J_{-2}^a,J_{-1}^bJ_{-1}^c,1} & = & \frac{1}{729}\Bigl[
30k\delta^{bc} T_1^a 
-128k\delta^{ac} T_3^b
-128k\delta^{ab} T_3^c 
+128k\delta^{ab} T_1^c
+128k\delta^{ac} T_1^b \nn\\
& & -384i kf^{abc} 
-296f^{bad}f^{cde} T_1^e 
-128 f^{bad}f^{cde} T_3^e 
+128 f^{bad}f^{cde} T_2^e \nn\\
& &
-24i f^{bad}T_1^dT_1^c 
-40i f^{bad}T_1^dT_3^c 
-40i f^{cad}T_1^dT_1^b \nn\\
& & +64i f^{cad}T_1^bT_3^d 
-64i f^{cad}T_1^bT_2^d  
+64i f^{bad}T_1^cT_3^d 
-64i f^{bad}T_1^cT_2^d \nn\\
& & -64i f^{bad}T_3^dT_3^c 
+64i f^{bad}T_2^dT_3^c 
+64i f^{bad} T_1^cT_1^d \nn\\
& & 
+30i f^{bcd}T_1^aT_2^d 
-40i f^{cad}T_1^dT_3^b 
-64i f^{cad}T_3^dT_3^b 
+64i f^{cad}T_2^dT_3^b \nn\\
& & 
-24 T_1^aT_1^cT_1^b
+24 T_1^aT_1^bT_3^c
+24 T_1^aT_1^cT_3^b 
-24 T_1^aT_3^cT_3^b\Bigr]
\eea
\bea
\vev{J_{-1}^a,J_{-2}^b,J_{-1}^c} & = & \frac{8}{729}\Bigl[
-12k\delta^{ac}T_2^b
-16k\delta^{bc}T_2^a
+16k\delta^{bc}T_3^a
-16k\delta^{ab} T_2^c 
+16k\delta^{ab} T_1^c \nn\\
& & 
+3if^{cad}T_2^bT_3^d
-3if^{cad}T_2^dT_2^b 
-16if^{bad} T_1^dT_2^c
+16if^{bad} T_1^cT_1^d \nn\\
& & 
-12if^{cad}T_1^dT_2^b
+16if^{bcd}T_3^aT_3^d
-16if^{bcd}T_2^aT_3^d \nn\\
& & 
-3T_1^cT_2^aT_2^b
+3T_1^cT_2^bT_3^a
+3T_2^aT_2^cT_2^b
-3T_2^cT_2^bT_3^a\Bigr]
\eea

Full expressions of $\vev{J_{-1}^a,J_{-1}^bJ_{-1}^c,J_{-1}^d}$ and 
$\vev{J_{-1}^aJ_{-1}^b,J_{-1}^cJ_{-1}^d,1}$ are very long. 
Therefore we write the expressions which are reduced to some of the 
above formulae.
\bea
\vev{J_{-1}^a,J_{-1}^bJ_{-1}^c,J_{-1}^d} & = &
\frac{16}{27}k\delta^{ab}\vev{1,J_{-1}^c,J_{-1}^d}
+\frac{16}{27}k\delta^{bd}\vev{J_{-1}^a,J_{-1}^c,1} \nn\\
& & -\frac{5}{27}k\delta^{bc}\vev{J_{-1}^a,1,J_{-1}^d} \nn\\
& & -\frac{2}{3\sqrt{3}}if^{bae}\vev{J_{-1}^e,J_{-1}^c,J_{-1}^d}
+\frac{2}{3\sqrt{3}}if^{bde}\vev{J_{-1}^a,J_{-1}^c,J_{-1}^e} \nn\\
& & +\frac{2}{3\sqrt{3}}\vev{J_{-1}^a,J_{-1}^c,J_{-1}^d}(T_3^b-T_1^b)
+\frac{16}{27}if^{bae}\vev{1,J_{-1}^c,J_{-1}^d}T_1^e \nn\\
& & +\frac{16}{27}if^{bde}\vev{J_{-1}^a,J_{-1}^c,1}T_3^e
-\frac{5}{27}if^{bce}\vev{J_{-1}^a,1,J_{-1}^d}T_2^e
\eea
\bea
\vev{J_{-1}^aJ_{-1}^b,J_{-1}^cJ_{-1}^d,1} & = &
-\frac{5}{27}k\delta^{cd}\vev{J_{-1}^aJ_{-1}^b,1,1}
+\frac{16}{27}if^{cae}f^{ebf}\vev{J_{-1}^f,J_{-1}^d,1} \nn\\
& & -\frac{2}{3\sqrt{3}}\vev{J_{-1}^aJ_{-1}^b,J_{-1}^d,1}(T_1^c+T_3^c) \nn\\
& & -\frac{2}{3\sqrt{3}}if^{cae}\vev{J_{-1}^eJ_{-1}^b,J_{-1}^d,1}
-\frac{2}{3\sqrt{3}}if^{cbe}\vev{J_{-1}^aJ_{-1}^e,J_{-1}^d,1} \nn\\
& & +\frac{16}{27}if^{cae}\vev{J_{-1}^b,J_{-1}^d,1}T_1^e
+\frac{16}{27}if^{cbe}\vev{J_{-1}^a,J_{-1}^d,1}T_1^e \nn\\
& & -\frac{5}{27}if^{cde}\vev{J_{-1}^aJ_{-1}^b,1,1}T_2^e
\eea

\section{The Action \label{appc}}
\setcounter{equation}{0}
In this appendix we collect the terms necessary for level 
[1,3], [2,6] and [5/2,5] calculations.
$S_0$ consists of the terms which survive after we take 
the Siegel gauge condition. 
$S_1$ consists of the linear terms in $\wt{t}, \wt{u}$ and $\wt{u}^a$,
which are excluded by Siegel gauge condition.
$c'=26-\frac{3k}{k+2}$ is the central charge of $X\times Y\times{\cal M}$.
\bea
S_0 & = & \Half(h_j-1)\vev{\Phi,\Phi}tt \nn\\
& & +\Half(h_j)\vev{J_{-1}^a\Phi,J_{-1}^b\Phi}u^au^b \nn\\
& & +\Half(h_j+1)\Biggl[\vev{J_{-2}^a\Phi,J_{-2}^b\Phi}v^av^b \nn\\
& & +2\vev{J_{-2}^a\Phi,J_{-1}^bJ_{-1}^c\Phi}v^av^{bc}
  +\vev{J_{-1}^aJ_{-1}^b\Phi,J_{-1}^cJ_{-1}^d\Phi}v^{ab}v^{cd}
  -\vev{\Phi,\Phi}\beta\beta\Biggr] \nn\\
& & +\frac{c'}{4}(h_j+1)\vev{\Phi,\Phi}ww \nn\\
& & +K^{3-h_{j_1}-h_{j_2}-h_{j_3}}
  \Biggl[\frac{1}{3}\vev{\Phi,\Phi,\Phi}ttt
   +\vev{\Phi,\Phi,J_{-1}^a\Phi}ttu^a
   +\vev{\Phi,J_{-1}^a\Phi,J_{-1}^b\Phi}tu^au^b \nn\\
& & +\frac{1}{3}\vev{J_{-1}^a\Phi,J_{-1}^b\Phi,J_{-1}^c\Phi}u^au^bu^c
   +\vev{\Phi,\Phi,J_{-2}^a\Phi}ttv^a
   +\vev{\Phi,\Phi,J_{-1}^aJ_{-1}^b\Phi}ttv^{ab}\nn\\
& & +\vev{\Phi,J_{-1}^a\Phi,J_{-2}^b\Phi}tu^av^b
   +\vev{\Phi,J_{-2}^a\Phi,J_{-1}^b\Phi}tv^au^b \nn\\
& & +\vev{\Phi,J_{-1}^a\Phi,J_{-1}^bJ_{-1}^c\Phi}tu^av^{bc}
   +\vev{\Phi,J_{-1}^aJ_{-1}^b\Phi,J_{-1}^c\Phi}tv^{ab}u^c \nn\\
& & +\vev{J_{-1}^a\Phi,J_{-1}^b\Phi,J_{-2}^c\Phi}u^au^bv^c
   +\vev{J_{-1}^a\Phi,J_{-1}^b\Phi,J_{-1}^cJ_{-1}^d\Phi}u^au^bv^{cd} \nn\\
& & +\vev{\Phi,J_{-2}^a\Phi,J_{-2}^b\Phi}tv^av^b \nn\\
& & +\vev{\Phi,J_{-2}^a\Phi,J_{-1}^bJ_{-1}^c\Phi}tv^av^{bc}
   +\vev{\Phi,J_{-1}^aJ_{-1}^b\Phi,J_{-2}^c\Phi}tv^{ab}v^c \nn\\
& & +\vev{\Phi,J_{-1}^aJ_{-1}^b\Phi,J_{-1}^cJ_{-1}^d\Phi}tv^{ab}v^{cd} \nn\\
& & -\frac{11}{27}\vev{\Phi,\Phi,J_{-2}^a\Phi}t\beta v^a
    -\frac{11}{27}\vev{\Phi,J_{-2}^a\Phi,\Phi}tv^a\beta \nn\\
& & -\frac{11}{27}\vev{\Phi,\Phi,J_{-1}^aJ_{-1}^b\Phi}t\beta v^{ab}
    -\frac{11}{27}\vev{\Phi,J_{-1}^aJ_{-1}^b\Phi,\Phi}tv^{ab}\beta \nn\\
& & -\frac{11}{27}\vev{\Phi,\Phi,\Phi}tt\beta
   -\frac{11}{27}\vev{\Phi,\Phi,\Phi}t\beta\beta \nn\\
& & -\frac{11}{27}\vev{\Phi,\Phi,J_{-1}^a\Phi}t\beta u^a
    -\frac{11}{27}\vev{\Phi,J_{-1}^a\Phi,\Phi}tu^a\beta \nn\\
& & -\frac{11}{27}\vev{J_{-1}^a\Phi,J_{-1}^b\Phi,\Phi}u^au^b\beta \nn\\
& & -\frac{5}{54}c'\vev{\Phi,\Phi,\Phi}ttw 
   +\left(\frac{128}{729}c'+(\frac{5}{54}c')^2\right)
   \vev{\Phi,\Phi,\Phi}tww \nn\\
& & -\frac{5}{54}c'\vev{\Phi,J_{-1}^a\Phi,\Phi}tu^aw
   -\frac{5}{54}c'\vev{\Phi,\Phi,J_{-1}^a\Phi}twu^a \nn\\
& & -\frac{5}{54}c'\vev{J_{-1}^a\Phi,J_{-1}^b\Phi,\Phi}u^au^bw \nn\\
& & +\frac{55}{1458}c'\vev{\Phi,\Phi,\Phi}t\beta w
   +\frac{55}{1458}c'\vev{\Phi,\Phi,\Phi}tw\beta \nn\\
& & -\frac{5}{54}c'\vev{\Phi,J_{-2}^a\Phi,\Phi}tv^aw
   -\frac{5}{54}c'\vev{\Phi,\Phi,J_{-2}^a\Phi}twv^a \nn\\
& & -\frac{5}{54}c'\vev{\Phi,J_{-1}^aJ_{-1}^b\Phi,\Phi}tv^{ab}w
   -\frac{5}{54}c'\vev{\Phi,\Phi,J_{-1}^aJ_{-1}^b\Phi}twv^{ab}\Biggr]
\eea
\bea
S_1 & = & \vev{\Phi,\Phi}\wt{t}(T^au^a) \nn\\
 & & -2\vev{\Phi,\Phi}\wt{u}(T^av^a)
 +2\vev{J_{-1}^a\Phi,J_{-1}^b\Phi}\wt{u}^av^b \nn\\
 & & -\vev{J_{-1}^a\Phi,J_{-1}^b\Phi}\wt{u}v^{ab}
 +\vev{J_{-1}^a\Phi,L_1J_{-1}^bJ_{-1}^c\Phi}\wt{u}^av^{bc} \nn\\
 & & -3\vev{\Phi,\Phi}\wt{u}\beta
 +\vev{\Phi,\Phi}\wt{u}^a(T^a\beta) \nn\\
 & & -\frac{c'}{2}\vev{\Phi,\Phi}\wt{u}w \nn\\
 & & +\frac{32}{81\sqrt{3}}\vev{\Phi,\Phi,\Phi}t\beta\wt{t}
 -\frac{32}{81\sqrt{3}}\vev{\Phi,\Phi,\Phi}\beta t\wt{t} \nn\\
 & & +\frac{16}{27}\vev{\Phi,\Phi,\Phi}tt\wt{u} \nn\\
 & & +\frac{16}{27}\vev{\Phi,J_{-1}^a\Phi,\Phi}tu^a\wt{u}
 +\frac{16}{27}\vev{J_{-1}^a\Phi,\Phi,\Phi}u^at\wt{u} \nn\\
 & & +\frac{16}{27}\vev{J_{-1}^a\Phi,J_{-1}^b\Phi,\Phi}u^au^b\wt{u} \nn\\
 & & +\frac{16}{27}\vev{\Phi,J_{-2}^a\Phi,\Phi}tv^a\wt{u}
 +\frac{16}{27}\vev{J_{-2}^a\Phi,\Phi,\Phi}v^at\wt{u} \nn\\
 & & +\frac{16}{27}\vev{\Phi,J_{-1}^aJ_{-1}^a\Phi,\Phi}tv^{ab}\wt{u}
 +\frac{16}{27}\vev{J_{-1}^aJ_{-1}^b\Phi,\Phi,\Phi}v^{ab}t\wt{u} \nn\\
 & & +\frac{32}{81\sqrt{3}}\vev{J_{-1}^a\Phi,\Phi,\Phi}u^a\beta\wt{t}
 -\frac{32}{81\sqrt{3}}\vev{\Phi,J_{-1}^a\Phi,\Phi}\beta u^a\wt{t} \nn\\
 & & +\frac{16}{81}\vev{\Phi,\Phi,\Phi}t\beta\wt{u}
 +\frac{16}{81}\vev{\Phi,\Phi,\Phi}t\wt{u}\beta \nn\\
 & & +\frac{32}{81\sqrt{3}}\vev{\Phi,\Phi,J_{-1}^a\Phi}t\beta\wt{u}^a
 -\frac{32}{81\sqrt{3}}\vev{\Phi,J_{-1}^a\Phi,\Phi}\beta t\wt{u}^a \nn\\
 & & -\frac{40}{729}c'\vev{\Phi,\Phi,\Phi}t\wt{u}w
   -\frac{40}{729}c'\vev{\Phi,\Phi,\Phi}tw\wt{u}\Biggr]
\eea

\section{Numerical data \label{appd}}
\setcounter{equation}{0}
\begin{center}

\begin{table}[htbp]
\label{tab1-1}
{\tiny
\begin{tabular}{|c|c|c|c|c|c|c|}\hline
 & $k=2$ & $k=3$ & $k=4$ & $k=5$ & $k=6$ & $k=10$ 
\\ \hline
$t^0$ & $0.148148$ & $0.185509$ & $0.200773$ & $0.209003$ & 
$0.214017$ & $0.222462$ 
\\ \hline 
$t^1$ & $0.320432$ & $0.365167$ & $0.381268$ & 
$0.388789$ & $0.392766$ & $0.397736$ 
\\ \hline 
$\sqrt{k}u^{01}$ & $0$ & $5.75964\times 10^{-17}$ & 
$-2.03876\times 10^{-17}$ & 
$-3.67281\times 10^{-17}$ & $-5.11186\times 10^{-17}$ & 
$4.42203\times 10^{-17}$ 
\\ \hline 
$\sqrt{k}u^{10}$ & $0$ & $-0.0369521$ & $-0.0484596$ & $-0.0543740$ & 
$-0.0575099$ & $-0.0596719$ 
\\ \hline 
$\sqrt{k}u^{11}$ & $0$ & $0$ & $0$ & $0$ & $0$ & $0$ 
\\ \hline 
$\sqrt{k}u^{12}$ & $0$ & $-0.0653227$ & $-0.0685322$ & $-0.0672844$ & 
$-0.0650648$ & $-0.0562591$ 
\\ \hline
$R^{[2,6]}_1$ & $0.579453$ & $0.601173$ & $0.614265$ & 
$0.624007$ & $0.631416$ & $0.648764$ 
\\ \hline
\end{tabular}
}

{\tiny
\begin{tabular}{|c|c|c|c|c|}\hline
 & $k=50$ & $k=1000$ & $k=100000$ & $k=100000000$ 
\\ \hline
$t^0$ & $0.227971$ & $0.228099$ & $0.228089$ & $0.228089$ 
\\ \hline 
$t^1$ & $0.39684$ & $0.395169$ & $0.395063$ & $0.395062$ 
\\ \hline 
$\sqrt{k}u^{01}$ & $7.93084\times 10^{-18}$ & 
$6.65144\times 10^{-16}$ & $2.52803\times 10^{-15}$ & 
$-6.73216\times 10^{-14}$ 
\\ \hline 
$\sqrt{k}u^{10}$ &
$-0.0382045$ & $-0.00932628$ & 
$-0.000936929$ & $-0.0000296296$ 
\\ \hline 
$\sqrt{k}u^{11}$ & $0$ & $0$ & $0$ & $0$ 
\\ \hline 
$\sqrt{k}u^{12}$ & 
$-0.0286686$ & $-0.00661447$ & 
$-0.000662529$ & $-0.0000209513$ 
\\ \hline
$R^{[2,6]}_1$ & $0.675916$ & $0.684152$ & $0.684611$ & $0.684616$ 
\\ \hline
\end{tabular}
}
\caption{${\cal T}^{[2,6]}_1$ and $R^{[2,6]}_1$ ($J=1/2$)}
\end{table}

\begin{table}[htbp]
\label{tab1-2-1}
{\tiny
\begin{tabular}{|c|c|c|c|c|c|c|c|c|c|c|c|}\hline
 & $k=4$ & $k=5$ & $k=6$ & $k=7$ & $k=8$ & $k=10$
\\ \hline
$t^0$ & $0.258107$ & $0.27399$ & $0.280901$ & $0.286079$ & 
$0.289726$ & $0.294389$ 
\\ \hline 
$t^1$ & $0.360168$ & $0.352909$ & $0.349867$ & $
  0.346261$ & $0.343223$ & $0.338634$ 
\\ \hline 
$t^2$ & $-0.158304$ & $-0.196357$ & $-0.210897$ & 
$-0.221076$ & $-0.227766$ & $-0.235527$ 
\\ \hline 
$\sqrt{k}u^{01}$ & $-9.72804\times 10^{-11}$ & 
$-9.55096\times 10^{-18}$ & $2.08702\times 10^{-17}$ & 
$1.83944\times 10^{-17}$ & $1.08632\times 10^{-18}$ & 
$1.40400\times 10^{-17}$ 
\\ \hline 
$\sqrt{k}u^{10}$ & $-2.45936\times 10^{-9}$ & $-0.0216722$ & 
$-0.0298519$ & $-0.0344495$ & $-0.0372043$ & $-0.0399092$ 
\\ \hline 
$\sqrt{k}u^{11}$ & $0$ & $0$ & $0$ & $0$ & $0$ & $0$
\\ \hline 
$\sqrt{k}u^{12}$ & $-3.45812\times 10^{-9}$ & $-0.0268181$ & 
$-0.0337738$ & $-0.0365391$ & $-0.0375822$ & $-0.0376270$ 
\\ \hline 
$\sqrt{k}u^{21}$ & $0$ & $0.0171184$ & $
  0.0249579$ & $0.0302724$ & $0.0340421$ & $0.0387376$ 
\\ \hline 
$\sqrt{k}u^{22}$ & $0$ & $0$ & $0$ & $0$ & $0$ & $0$
\\ \hline 
$\sqrt{k}u^{23}$ & $0$ & $0.0372721$ & $0.0458503$ & $0.0494348$ & $
  0.0509581$ & $0.0513974$ 
\\ \hline
$R^{[2,6]}_1$ & $0.67064$ & $0.656018$ & $0.648754$ & $0.648206$ & 
$0.649353$ & $0.652857$ 
\\ \hline
\end{tabular}
}

{\tiny
\begin{tabular}{|c|c|c|c|c|}\hline
 & $k=50$ & $k=1000$ & $k=100000$ & $k=100000000$ 
\\ \hline
$t^0$ & $0.303815$ & $0.304131$ & $0.304119$ & $0.304119$ 
\\ \hline 
$t^1$ & $0.324627$ & $0.322641$ & $0.322567$ & $0.322567$ 
\\ \hline 
$t^2$ & $-0.243488$ & $-0.240632$ & $-0.240429$ & $-0.240427$ 
\\ \hline 
$\sqrt{k}u^{01}$ & 
$-4.04713\times 10^{-17}$ & $-8.12139\times 10^{-17}$ & 
$1.79546\times 10^{-15}$ & $-1.37752\times 10^{-16}$ 
\\ \hline 
$\sqrt{k}u^{10}$ & 
$-0.0299926$ & $-0.00759889$ & $-0.000764983$ & $-0.0000241925$ 
\\ \hline 
$\sqrt{k}u^{11}$ & $0$ & $0$ & $0$ & $0$ 
\\ \hline 
$\sqrt{k}u^{12}$ & 
$-0.0225064$ & $-0.00538937$ & $-0.000540939$ & $-0.0000171067$ 
\\ \hline 
$\sqrt{k}u^{21}$ & 
$0.0351798$ & $0.00928736$ & $0.000936888$ & $0.0000296296$ 
\\ \hline 
$\sqrt{k}u^{22}$ & $0$ & $0$ & $0$ & $0$ 
\\ \hline 
$\sqrt{k}u^{23}$ & 
$0.0317163$ & $0.00762109$ & $0.000765006$ & $0.0000241925$ 
\\ \hline
$R^{[2,6]}_1$ & $0.676028$ & $0.684152$ & $0.684611$ & $0.684616$
\\ \hline
\end{tabular}
}
\caption{${\cal T}^{[2,6]}_1$ and $R^{[2,6]}_1$ ($J=1$)}
\end{table}

\begin{table}[htbp]
\label{tab1-2-2}
{\tiny
\begin{tabular}{|c|c|c|c|c|c|c|}\hline
 & $k=4$ & $k=5$ & $k=6$ & $k=7$ & $k=8$ & $k=10$ 
\\ \hline
$t^0$ &     & $0.0816382$ & $0.102873$ & $0.115138 $ & $0.123276$ & 
$0.133189$ 
\\ \hline 
$t^1$ &     & $0.161194$ & $0.230157$ & $0.258811$ & $0.27566$ & 
$0.29424$ 
\\ \hline 
$t^2$ &     & $0.288496$ & $0.283243$ & $0.279293$ & $0.2752$ & 
$0.268238$ 
\\ \hline 
$\sqrt{k}u^{01}$ &     & $-2.92522\times 10^{-17}$ & 
$1.52745\times 10^{-17}$ & 
$3.35243\times 10^{-18}$ & $-3.21793\times 10^{-18}$ & 
$-1.54667\times 10^{-17}$ 
\\ \hline 
$\sqrt{k}u^{10}$ &     & $-0.0906831$ & $-0.0861704$ & $-0.0844246$ & 
$-0.0827142$ & $-0.0794370$ 
\\ \hline 
$\sqrt{k}u^{11}$ &     & $0$ & $0$ & $0$ & $0$ & $0$ 
\\ \hline 
$\sqrt{k}u^{12}$ &     & $-0.112214$ & $-0.0974907$ & $-0.0895457$ & 
$-0.0835540$ & $-0.0748938$ 
\\ \hline 
$\sqrt{k}u^{21}$ &     & $-0.00399811$ & $-0.00923793$ & $-0.0125176$ & 
$-0.0147493$ & $-0.0174074$ 
\\ \hline 
$\sqrt{k}u^{22}$ &     & $0$ & $0$ & $0$ & $0$ & $0$ 
\\ \hline 
$\sqrt{k}u^{23}$ &     & $-0.00870517$ & $-0.0169712$ & $-0.0204412$ & 
$-0.0220783$ & $-0.0230962$ 
\\ \hline
$R^{[2,6]}_2$ &     & $0.460292$ & $0.563618$ & 
$0.589081$ & $0.603074$ & $0.619284$ 
\\ \hline
\end{tabular}
}

{\tiny
\begin{tabular}{|c|c|c|c|c|}\hline
 & $k=50$ & $k=1000$ & $k=100000$ & $k=100000000$ 
\\ \hline
$t^0$ & $0.151627$ & $0.152091$ & $0.15206$ & $0.152059$ 
\\ \hline 
$t^1$ & $0.322597$ & $0.322644$ & $0.322567$ & $0.322567$ 
\\ \hline 
$t^2$ & $0.244549$ & $0.240599$ & $0.240429$ & $0.240427$ 
\\ \hline 
$\sqrt{k}u^{01}$ & 
$4.16731\times 10^{-17}$ & $7.71083\times 10^{-16}$ & 
$-2.70688\times 10^{-15}$ & $-1.03728\times 10^{-15}$ 
\\ \hline 
$\sqrt{k}u^{10}$ & 
$-0.0475205$ & $-0.0115869$ & $-0.00116351$ & $-0.0000570216$ 
\\ \hline 
$\sqrt{k}u^{11}$ & $0$ & $0$ & $0$ & $0$ 
\\ \hline 
$\sqrt{k}u^{12}$ & 
$-0.0356593$ & $-0.00821778$ & 
$-0.000822749$ & $-0.0000403204$ 
\\ \hline 
$\sqrt{k}u^{21}$ & 
$-0.0165016$ & $-0.00443557$ & 
$-0.000448828$ & $0.0000105777$ 
\\ \hline 
$\sqrt{k}u^{22}$ & $0$ & $0$ & $0$ & $0$ 
\\ \hline 
$\sqrt{k}u^{23}$ & 
$-0.0148770$ & $-0.00363978$ & 
$-0.000366486$ & $8.63666\times 10^{-6}$ 
\\ \hline
$R^{[2,6]}_2$ & $0.668068$ & $0.683691$ & $0.684607$ & $0.684616$ 
\\ \hline
\end{tabular}
}
\caption{${\cal T}^{[2,6]}_2$ and $R^{[2,6]}_2$ ($J=1$)}
\end{table}

\begin{table}[htbp]
\label{tab2-1}
{\tiny
\begin{tabular}{|c|c|c|c|c|c|c|c|c|c|}\hline
 & $k=2$ & $k=3$ & $k=4$ & $k=5$ & $k=6$ & $k=7$ & $k=8$ & $k=9$ & 
$k=10$ \\ \hline
$t^0$ & $0.182129$ & $0.225759$ & $0.242631$ & $0.251724$ & $0.25716$ & 
$0.260658$ & $0.263032$ & $0.26471$ & $0.265934$ \\ \hline 
$t^1$ & $0.36068$ & $0.426136$ & $0.445349$ & $0.454849$ & $0.460191$ & 
$0.46345$ & $0.465557$ & $0.466977$ & $0.467967$ \\ \hline 
$\sqrt{k}u^{0,1}$ & $0$ & $0.0124587$ & $0.00358094$ & $0.00187705$ & 
$0.00120394$ & $0.00085194$ & $0.000638823$ & $0.000497805$ & $0.000399083$
\\ \hline 
$\sqrt{k}u^{1,0}$ & $0$ & $-0.0362956$ & $-0.057079$ & $-0.0655588$ & 
$-0.0697306$ & $-0.0717213$ & $-0.0724835$ & $-0.0725181$ & $-0.0721097$
\\ \hline 
$\sqrt{k}u^{1,1}$ & $0$ & $0$ & $0$ & $0$ & $0$ & $0$ & $0$ & $0$ & 
$0$ \\ \hline 
$\sqrt{k}u^{1,2}$ & $0$ & $-0.0934329$ & $-0.0890626$ & $-0.085463$ & 
$-0.0816555$ & $-0.0780169$ & $-0.0746705$ & $-0.0716331$ & $-0.0688848$
\\ \hline 
$\sqrt{k}v^{0,1}$ & $0$ & $0$ & $0$ & $0$ & $0$ & $0$ & $0$ & $0$ & $0$
\\ \hline 
$\sqrt{k}v^{1,0}$ &     & $0$ & $0$ & $0$ & $0$ & $0$ & $0$ & $0$ & $0$
\\ \hline 
$\sqrt{k}v^{1,1}$ &     & $-0.0416215$ & $-0.0373648$ & $-0.0378112$ & 
$-0.0383651$ & $-0.0386618$ & $-0.0387178$ & $-0.0385935$ & $-0.0383426$
\\ \hline 
$\sqrt{k}v^{1,2}$ &     & $0$ & $0$ & $0$ & $0$ & $0$ & $0$ & $0$ & $0$
 \\ \hline 
$kv^{0,1,0}$ & $-0.0032832$ & $-0.00588747$ & $-0.0101722$ & $-0.012695$ & 
$-0.0144098$ & $-0.0156477$ & $-0.016579$ & $-0.0173019$ & 
$-0.0178773$ \\ \hline 
$kv^{0,1,1}$ & $0$ & $0$ & $0$ & $0$ & $0$ & $0$ & $0$ & $0$ & $0$
\\ \hline 
$kv^{0,1,2}$ & $0.015965$ & $0.0106885$ & $0.00825364$ & $0.0066016$ & 
$0.00549079$ & $0.00469881$ & $0.00410654$ & $0.003647$ & 
$0.00327999$ \\ \hline 
$kv^{1,0,1}$ &     & $-0.048801$ & $-0.0385105$ & $-0.0395883$ & 
$-0.0422954$ & $-0.0452372$ & $-0.0480562$ & $-0.0506541$ & $-0.0530145$
\\ \hline 
$kv^{1,1,0}$ &     & $0$ & $0$ & $0$ & $0$ & $0$ & $0$ & $0$ & $0$
\\ \hline 
$kv^{1,1,1}$ &     & $0.0485517$ & $0.038505$ & $0.0393737$ & $0.0418666$ & 
$0.0446333$ & $0.0473165$ & $0.0498103$ & $0.0520912$ 
\\ \hline 
$kv^{1,1,2}$ &     & $0$ & $0$ & $0$ & $0$ & $0$ & $0$ & $0$ & $0$
\\ \hline 
$kv^{1,2,1}$ &     & $0.0312945$ & $0.0247845$ & $0.025381$ & $0.0270238$ & 
$0.0288359$ & $0.0305873$ & $0.032211$ & $0.0336933$
\\ \hline 
$kv^{1,2,2}$ &     & $0$ & $0$ & $0$ & $0$ & $0$ & $0$ & $0$ & $0$
\\ \hline 
$kv^{1,2,3}$ &     & $0.00641082$ & $0.0067976$ & $0.0065752$ & 
$0.00613662$ & $0.00567264$ & $0.00523985$ & $0.0048518$ & $0.00450837$
\\ \hline 
$\beta^0$ & $-0.06644$ & $-0.0776835$ & $-0.0825814$ & $-0.085021$ & 
$-0.0863635$ & $-0.0871489$ & $-0.0876249$ & $-0.0879181$ & 
$-0.0880984$ \\ \hline 
$\beta^1$ &     & $-0.0609899$ & $-0.0736911$ & $-0.083145$ & $-0.0904823$ & 
$-0.0963519$ & $-0.101159$ & $-0.10517$ & $-0.108569$ 
\\ \hline 
$w^0$ & $0.0252397$ & $0.0263729$ & $0.0274827$ & $0.0279216$ & 
$0.0280881$ & $0.02813$ & $0.028111$ & $0.0280617$ & $0.0279978$ 
\\ \hline 
$w^1$ &     & $0.0186859$ & $0.0223821$ & $0.025152$ & $0.0273088$ & 
$0.0290367$ & $0.0304532$ & $0.0316362$ & $0.0326396$ 
\\ \hline
$R^{[5/2,5]}_1$ & $0.751672$ & $0.821284$ & $0.838754$ & 
$0.853024$ & $0.864079$ & 
$0.872784$ & $0.879797$ & $0.885567$ & $0.890403$
\\ \hline 
\end{tabular}
}

{\tiny
\begin{tabular}{|c|c|c|c|c|}\hline
 & $k=50$ & $k=1000$ & $k=100000$ & $k=100000000$ 
\\ \hline
$t^0$ & $0.27083$ & $0.270809$ & $0.270796$ & $0.270795$ 
\\ \hline 
$t^1$ & $0.470276$ & $0.469119$ & $0.469032$ & $0.469031$ 
\\ \hline 
$\sqrt{k}u^{0,1}$ & 
$0.0000116695$ & $7.84482\times 10^{-9}$ & $6.93279\times 10^{-14}$ & 
$4.38356\times 10^{-7}$ 
\\ \hline 
$\sqrt{k}u^{1,0}$ & 
$-0.0445576$ & $-0.0106754$ & $-0.00107121$ & $-0.0000334374$ 
\\ \hline 
$\sqrt{k}u^{1,1}$ & 
$0$ & $0$ & $0$ & $0$ 
\\ \hline 
$\sqrt{k}u^{1,2}$ & 
$-0.0334611$ & $-0.00757131$ & $-0.000757483$ & $-0.0000245737$ 
\\ \hline 
$\sqrt{k}v^{0,1}$ & $0$ & $0$ & $0$ & $0$ 
\\ \hline 
$\sqrt{k}v^{1,0}$ & $0$ & $0$ & $0$ & $0$ 
\\ \hline 
$\sqrt{k}v^{1,1}$ & $-0.0249031$ & $-0.00614266$ & 
$-0.00061746$ & $-0.0000195268$ 
\\ \hline 
$\sqrt{k}v^{1,2}$ & $0$ & $0$ & $0$ & $0$ 
\\ \hline 
$kv^{0,1,0}$ & $-0.0217152$ & $-0.0224384$ & $-0.0224725$ & 
$-0.0224728$ 
\\ \hline 
$kv^{0,1,1}$ & $0$ & $0$ & $0$ & $0$ 
\\ \hline 
$kv^{0,1,2}$ & $0.00064918$ & $0.0000321639$ & $3.21426\times 10^{-7}$ & 
$3.35187\times 10^{-10}$ 
\\ \hline 
$kv^{1,0,1}$ & $-0.0801125$ & $-0.0902368$ & $-0.0908171$ & 
$-0.0908230$ 
\\ \hline 
$kv^{1,1,0}$ & $0$ & $0$ & $0$ & $0$ 
\\ \hline 
$kv^{1,1,1}$ & $0.078991$ & $0.0892917$ & $0.0898854$ & $0.0898914$ 
\\ \hline 
$kv^{1,1,2}$ & $0$ & $0$ & $0$ & $0$ 
\\ \hline 
$kv^{1,2,1}$ & $0.0510425$ & $0.0576411$ & $0.0580209$ & $0.580248$ 
\\ \hline 
$kv^{1,2,2}$ & $0$ & $0$ & $0$ & $0$ 
\\ \hline 
$kv^{1,2,3}$ & $0.00111353$ & $0.0000581099$ & $5.82313\times 10^{-7}$ & 
$5.91053\times 10^{-10}$ 
\\ \hline 
$\beta^0$ & $-0.0874818$ & $-0.0866824$ & $-0.0866323$ & 
$-0.0866318$ 
\\ \hline 
$\beta^1$ & $-0.139739$ & $-0.149504$ & $-0.150045$ & $-0.150051$ 
\\ \hline 
$w^0$ & $0.0266379$ & $0.0259881$ & $0.0259497$ & $0.0259494$ 
\\ \hline 
$w^1$ & $0.0418775$ & $0.0447829$ & $0.044944$ & $0.0449456$ 
\\ \hline
$R^{[5/2,5]}_1$ & $0.93371$ & $0.947746$ & $0.948545$ & $0.948553$
\\ \hline
\end{tabular}
}
\caption{${\cal T}^{[5/2,5]}_1$ and $R^{[5/2,5]}_1$ ($J=1/2$)}
\end{table}

\clearpage

{\tiny
\begin{tabular}{|c|c|c|c|c|c|c|c|c|c|c|c|c|c|c|c|c|c|c|}\hline
 & $k=4$ & $k=5$ & $k=6$ & $k=7$ & $k=8$ & $k=9$ & $k=10$ 
\\ \hline
$t^0$ & $0.314788$ & $0.331704$ & $0.338216$ & $0.343281$ & 
$0.346835$ & $0.349408$ & $0.351321$ 
\\ \hline 
$t^1$ & $0.41735$ & $0.409288$ & $0.407187$ & $0.404102$ & 
$0.401494$ & $0.399346$ & $0.397583$ 
\\ \hline 
$t^2$ & $-0.184485$ & $-0.225581$ & $-0.239889$ & $-0.249824$ & 
$-0.256177$ & $-0.260388$ & $-0.263254$ 
\\ \hline 
$\sqrt{k}u^{0,1}$ & $1.02888\times 10^{-15}$ & $0.000889899$ & 
$0.000605516$ & $0.000407051$ & $0.000286384$ & $0.000210621$ & $0.000160798$ 
\\ \hline 
$\sqrt{k}u^{1,0}$ & $1.08749\times 10^{-15}$ & $-0.0252814$ & 
$-0.0350701$ & $-0.0406004$ & $-0.0439127$ & $-0.0459348$ & $-0.0471537$ 
\\ \hline 
$\sqrt{k}u^{1,1}$ & $0$ & $0$ & $0$ & $0$ & $0$ & $0$ & $0$ 
\\ \hline 
$\sqrt{k}u^{1,2}$ & $-1.55042\times 10^{-15}$ & $-0.0338029$ & 
$-0.0413802$ & $-0.0442018$ & $-0.0451556$ & $-0.0452445$ & $-0.0449002$ 
\\ \hline 
$\sqrt{k}u^{2,1}$ & $-5.77304\times 10^{-16}$ & $0.0211377$ & 
$0.0308004$ & $0.0372146$ & $0.0416885$ & $0.0448659$ & 
$0.0471382$ 
\\ \hline 
$\sqrt{k}u^{2,2}$ & $0$ & $0$ & $0$ & $0$ & $0$ & $0$ & $0$ 
\\ \hline 
$\sqrt{k}u^{2,3}$ & $7.13522\times 10^{-16}$ & $0.0479757$ & 
$0.0578942$ & $0.0616537$ & $0.0630256$ & $0.0632565$ & $0.0628917$ 
\\ \hline 
$\sqrt{k}v^{0,1}$ & $0$ & $0$ & $0$ & $0$ & $0$ & $0$ & $0$ \\ \hline 
$\sqrt{k}v^{1,0}$ & $0$ & $0$ & $0$ & $0$ & $0$ & $0$ & $0$ \\ \hline 
$\sqrt{k}v^{1,1}$ & $-0.0417624$ & $-0.0362672$ & $-0.0336249$ & 
$-0.0319554$ & $-0.0307608$ & $-0.0298178$ & $-0.0290239$ 
\\ \hline 
$\sqrt{k}v^{1,2}$ & $0$ & $0$ & $0$ & $0$ & $0$ & $0$ & $0$ \\ \hline 
$\sqrt{k}v^{2,1}$ &     &     &     &     &     &     &     \\ \hline 
$\sqrt{k}v^{2,2}$ &     &     &     &     &     &     &     \\ \hline 
$\sqrt{k}v^{2,3}$ &     &     &     &     &     &     &     \\ \hline 
$kv^{0,1,0}$ & $-0.01387752$ & $-0.0172971$ & $-0.0194408$ & 
$-0.021064$ & $-0.022311$ & $-0.0232982$ & $-0.0240986$ 
\\ \hline 
$kv^{0,1,1}$ & $0$ & $0$ & $0$ & $0$ & $0$ & $0$ & $0$ 
\\ \hline 
$kv^{0,1,2}$ & $0.00899444$ & $0.00757175$ & $0.00645066$ & 
$0.00563377$ & $0.00500638$ & $0.00450982$ & $0.00410681$ 
\\ \hline 
$kv^{1,0,1}$ & $-0.068786$ & $-0.0607325$ & $-0.0568655$ & 
$-0.0548976$ & $-0.0539287$ & $-0.0534971$ & $-0.0533688$ 
\\ \hline 
$kv^{1,1,0}$ & $0$ & $0$ & $0$ & $0$ & $0$ & $0$ & $0$ 
\\ \hline 
$kv^{1,1,1}$ & $0.069336$ & $0.0623375$ & $0.0586612$ & 
$0.0566855$ & $0.0556311$ & $0.0550894$ & $0.0548466$ 
\\ \hline 
$kv^{1,1,2}$ & $0$ & $0$ & $0$ & $0$ & $0$ & $0$ & $0$ 
\\ \hline 
$kv^{1,2,1}$ & $0.0445292$ & $0.0398361$ & $0.0374356$ & $0.0361654$ & 
$0.0355021$ & $0.0351731$ & $0.0350373$ 
\\ \hline 
$kv^{1,2,2}$ & $0$ & $0$ & $0$ & $0$ & $0$ & $0$ & $0$ 
\\ \hline 
$kv^{1,2,3}$ & $-0.0123896$ & $-0.0088196$ & $-0.00622704$ & 
$-0.00462895$ & $-0.00357675$ & $-0.0028504$ & $-0.00232872$ 
\\ \hline 
$kv^{2,1,0}$ &     &     &     &     &     &     &     
\\ \hline 
$kv^{2,1,1}$ &     &     &     &     &     &     &     
\\ \hline 
$kv^{2,1,2}$ &     &     &     &     &     &     &      
\\ \hline 
$kv^{2,2,1}$ &     &     &     &     &     &     &      
\\ \hline 
$kv^{2,2,2}$ &     &     &     &     &     &     &     
\\ \hline 
$kv^{2,2,3}$ &     &     &     &     &     &     &    
\\ \hline 
$kv^{2,3,2}$ &     &     &     &     &     &     &     
\\ \hline 
$kv^{2,3,3}$ &     &     &     &     &     &     &    
\\ \hline 
$kv^{2,3,4}$ &     &     &     &     &     &     &     
\\ \hline 
$\beta^0$ & $-0.105789$ & $-0.110894$ & $-0.112718$ & $-0.114132$ & 
$-0.115101$ & $-0.115783$ & $-0.116275$ 
\\ \hline 
$\beta^1$ & $-0.0669474$ & $-0.0721441$ & $-0.0770841$ & $-0.0807792$ & 
$-0.0837999$ & $-0.0863316$ & $-0.0884924$
\\ \hline 
$\beta^2$ &     &     &     &     &     &     &     
\\ \hline 
$w^0$ & $0.0346485$ & $0.0359941$ & $0.0363952$ & $0.0367119$ & 
$0.0369176$ & $0.0370535$ & $0.0371441$ 
\\ \hline 
$w^1$ & $0.0188242$ & $0.0199966$ & $0.0212771$ & $0.0222087$ & 
$0.0229717$ & $0.0236128$ & $0.0241617$ 
\\ \hline 
$w^2$ &     &     &     &     &     &     &     
\\ \hline
$R^{[5/2,5]}_1$ & $0.933395$ & $0.908232$ & $0.894435$ & 
$0.891439$ & $0.891457$ & 
$0.892623$ & $0.894206$
\\ \hline
\end{tabular}
}

\begin{table}[htbp]
\label{tab2-2-1}
{\tiny
\begin{tabular}{|c|c|c|c|c|c|c|c|c|c|c|c|c|c|c|c|c|c|c|}\hline
 & $k=12$ & $k=14$ & $k=17$ &$k=50$ & 
$k=1000$ & $k=100000$ & $k=100000000$ 
\\ \hline
$t^0$ & 
$0.355306$ & $0.357005$ & $0.358529$ & 
$0.361078$ & $0.361079$ & $0.361061$ & $0.361061$ 
\\ \hline 
$t^1$ & 
$0.393787$ & $0.391735$ & $0.389663$ & $0.384585$ & 
$0.383032$ & $0.382963$ & $0.382962$ 
\\ \hline 
$t^2$ & 
$-0.278162$ & $-0.281215$ & $-0.283798$ & $-0.286995$ & 
$-0.285579$ & $-0.285445$ & $-0.285443$ 
\\ \hline 
$\sqrt{k}u^{0,1}$ & 
$0.000138589$ & $0.000111494$ & $0.0000830053$ & $0.0000109995$ & 
$9.38782\times 10^{-9}$ & $1.00141\times 10^{-13}$ & 
$2.95267\times 10^{-8}$
\\ \hline 
$\sqrt{k}u^{1,0}$ & 
$-0.0493389$ & $-0.0493019$ & $-0.0483803$ & $-0.0353295$ & 
$-0.00870582$ & $-0.000874628$ & $-0.0000276233$
\\ \hline 
$\sqrt{k}u^{1,1}$ & 
$0$ & $0$ & $0$ & $0$ & $0$ & $0$ & $0$ 
\\ \hline 
$\sqrt{k}u^{1,2}$ & 
$-0.0447839$ & $-0.0432117$ & $-0.0408498$ & 
$-0.0265403$ & $-0.00617445$ & $-0.000618474$ & $-0.0000196093$
\\ \hline 
$\sqrt{k}u^{2,1}$ & 
$0.0515573$ & $0.0528416$ & $0.0532301$ & $0.0418617$ & 
$0.0106457$ & $0.00107118$ & $0.0000338463$
\\ \hline 
$\sqrt{k}u^{2,2}$ & 
$0$ & $0$ & $0$ & $0$ & $0$ & $0$ & $0$ 
\\ \hline 
$\sqrt{k}u^{2,3}$ & 
$0.0634499$ & $0.0613673$ & $0.0581311$ & $0.0377635$ & 
$0.00873577$ & $0.000874658$ & $0.0000276956$
\\ \hline 
$\sqrt{k}v^{0,1}$ & 
$0$ & $0$ & $0$ & $0$ & $0$ & $0$ & $0$ 
\\ \hline 
$\sqrt{k}v^{1,0}$ & 
$0$ & $0$ & $0$ & $0$ & $0$ & $0$ & $0$
\\ \hline 
$\sqrt{k}v^{1,1}$ & 
$-0.0304986$ & $-0.0295192$ & $-0.0282152$ & $-0.019975$ & 
$-0.00500974$ & $-0.000504147$ & $-0.0000159482$
\\ \hline 
$\sqrt{k}v^{1,2}$ & $0$ & $0$ & $0$ & $0$ & 
$0$ & $0$ & $0$ 
\\ \hline 
$\sqrt{k}v^{2,1}$ & $0$ & $0$ & $0$ & $0$ & 
$0$ & $0$ & $0$ 
\\ \hline 
$\sqrt{k}v^{2,2}$ & 
$0.0197379$ & $0.0215403$ & $0.0231784$ & $0.0224966$ & 
$0.00642606$ & $0.000650809$ & $0.0000205775$
\\ \hline 
$\sqrt{k}v^{2,3}$ & $0$ & $0$ & $0$ & $0$ & 
$0$ & $0$ & $0$ 
\\ \hline 
$kv^{0,1,0}$ & 
$-0.0247824$ & $-0.0255961$ & $-0.0264488$ & $-0.0289154$ & 
$-0.0299178$ & $-0.0299633$ & $-0.0299637$
\\ \hline 
$kv^{0,1,1}$ & $0$ & $0$ & $0$ & $0$ & $0$ & 
$0$ & $0$ 
\\ \hline 
$kv^{0,1,2}$ & 
$0.00345395$ & $0.00298358$ & $0.00247767$ & $0.000860433$ & 
$0.0000428906$ & $4.28569\times 10^{-7}$ & $4.28666\times 10^{-10}$
\\ \hline 
$kv^{1,0,1}$ & 
$-0.0561326$ & $-0.0570219$ & $-0.058401$ & $-0.0665359$ & 
$-0.0736984$ & $-0.0741519$ & $-0.0741946$
\\ \hline 
$kv^{1,1,0}$ & $0$ & $0$ & $0$ & $0$ & $0$ & 
$0$ & $0$ 
\\ \hline 
$kv^{1,1,1}$ & 
$0.0574638$ & $0.0581291$ & $0.0592386$ & 
$0.0663568$ & $0.0729662$ & $0.0733916$ & $0.0734289$
\\ \hline 
$kv^{1,1,2}$ & $0$ & $0$ & $0$ & $0$ & 
$0$ & $0$ & $0$ 
\\ \hline 
$kv^{1,2,1}$ & 
$0.0367481$ & $0.0372163$ & $0.0379787$ & $0.0427422$ & 
$0.0470952$ & $0.0473741$ & $0.0474194$
\\ \hline 
$kv^{1,2,2}$ & $0$ & $0$ & $0$ & $0$ & $0$ & 
$0$ & $0$ 
\\ \hline 
$kv^{1,2,3}$ & 
$-0.00117633$ & $-0.000764075$ & $-0.000410992$ & $0.000105204$ & 
$0.0000119647$ & $1.23327\times 10^{-7}$ & $1.25443\times 10^{-10}$
\\ \hline 
$kv^{2,1,0}$ & 
$-0.00106331$ & $-0.00127834$ & $-0.00143850$ & $-0.00108355$ & 
$-0.0000752970$ & $-7.65221\times 10^{-7}$ & 
$-7.62379\times 10^{-10}$
\\ \hline 
$kv^{2,1,1}$ & $0$ & $0$ & $0$ & $0$ & 
$0$ & $0$ & $0$ 
\\ \hline 
$kv^{2,1,2}$ & 
$0.0185435$ & $0.0249567$ & $0.0332872$ & $0.0730424$ & 
$0.104021$ & $0.105919$ & $0.105885$
\\ \hline 
$kv^{2,2,1}$ & $0$ & $0$ & $0$ & $0$ & 
$0$ & $0$ & $0$ 
\\ \hline 
$kv^{2,2,2}$ & 
$-0.00888098$ & $-0.0122799$ & $-0.0166614$ & 
$-0.0374381$ & $-0.0536953$ & $-0.0546959$ & $-0.0546833$
\\ \hline 
$kv^{2,2,3}$ & $0$ & $0$ & $0$ & $0$ & 
$0$ & $0$ & $0$ 
\\ \hline 
$kv^{2,3,2}$ & 
$-0.0137015$ & $-0.0186574$ & $-0.0250729$ & 
$-0.0556006$ & $-0.0794333$ & $-0.0808965$ & $-0.0808575$
\\ \hline 
$kv^{2,3,3}$ & $0$ & $0$ & $0$ & $0$ & 
$0$ & $0$ & $0$ 
\\ \hline 
$kv^{2,3,4}$ & 
$-0.00632142$ & $-0.00589583$ & $-0.00527583$ & $-0.00223821$ & 
$-0.000122223$ & $-1.22723\times 10^{-6}$ & $-1.23027\times 10^{-9}$
\\ \hline 
$\beta^0$ & 
$-0.116649$ & $-0.116923$ & $-0.11709$ & $-0.116564$ & 
$-0.115576$ & $-0.115510$ & $-0.115509$ 
\\ \hline 
$\beta^1$ & 
$-0.0934840$ & $-0.0964424$ & $-0.0998803$ & 
$-0.113246$ & $-0.122002$ & $-0.122511$ & $-0.122516$ 
\\ \hline 
$\beta^2$ & 
$0.0378906$ & $0.043247$ & $0.0495507$ & $0.0742366$ & 
$0.0903689$ & $0.0913084$ & $0.0913180$ 
\\ \hline 
$w^0$ & 
$0.0365577$ & $0.0364844$ & $0.0363524$ & $0.0354623$ & 
$0.0346507$ & $0.0345997$ & $0.0345991$ 
\\ \hline 
$w^1$ &
$0.0267857$ & $0.0277387$ & $0.0288614$ & $0.0333927$ & 
$0.0365112$ & $0.0366961$ & $0.0366979$ 
\\ \hline 
$w^2$ & 
$-0.0119448$ & $-0.0135931$ & $-0.0154986$ & $-0.0226491$ & 
$-0.0270963$ & $-0.0273504$ & $-0.0273530$ 
\\ \hline
$R^{[5/2,5]}_1$ & $0.904861$ & $0.908966$ & $0.913995$ & $0.934266$ & 
$0.947748$ & $0.948545$ & $0.948553$
\\ \hline
\end{tabular}
}
\caption{${\cal T}^{[5/2,5]}_1$ and $R^{[5/2,5]}_1$ ($J=1$)}
\end{table}

{\tiny
\begin{tabular}{|c|c|c|c|c|c|c|c|c|c|c|c|c|c|c|c|c|c|c|}\hline
 & $k=4$ & $k=5$ & $k=6$ & $k=7$ & $k=8$ & $k=9$ & $k=10$ 
\\ \hline
$t^0$ &     & $0.108574$ & $0.130853$ & $0.143935$ & $0.152574$ & 
$0.158614$ & $0.163012$ 
\\ \hline 
$t^1$ &     & $0.216655$ & $0.282623$ & $0.311856$ & $0.329479$ & 
$0.341154$ & $0.34935$ 
\\ \hline 
$t^2$ &     & $0.306499$ & $0.30633$ & $0.305413$ & $0.303459$ & 
$0.301319$ & $0.299306$ 
\\ \hline 
$\sqrt{k}u^{0,1}$ &     & $-0.0000180229$ & $0.000129681$ & $0.000184364$ & 
$0.000191436$ & $0.000180074$ & $0.000162688$ 
\\ \hline 
$\sqrt{k}u^{1,0}$ &     & $-0.114929$ & $-0.108674$ & $-0.106245$ & 
$-0.103925$ & $-0.101672$ & $-0.0994938$ 
\\ \hline 
$\sqrt{k}u^{1,1}$ &     & $0$ & $0$ & $0$ & $0$ & $0$ & $0$ 
\\ \hline 
$\sqrt{k}u^{1,2}$ &     & $-0.142204$ & $-0.123077$ & $-0.112891$ & 
$-0.105201$ & $-0.0990687$ & $-0.0940038$ 
\\ \hline 
$\sqrt{k}u^{2,1}$ &     & $-0.00429864$ & $-0.0104174$ & $-0.0143277$ & 
$-0.0170257$ & $-0.0189278$ & $-0.0202891$ 
\\ \hline 
$\sqrt{k}u^{2,2}$ &     & $0$ & $0$ & $0$ & $0$ & $0$ & $0$ 
\\ \hline 
$\sqrt{k}u^{2,3}$ &     & $-0.00933469$ & $-0.0193081$ & $-0.0236304$ & 
$-0.0257211$ & $-0.0267095$ & $-0.0271114$ 
\\ \hline 
$\sqrt{k}v^{0,1}$ &     & $0$ & $0$ & $0$ & $0$ & $0$ & $0$ 
\\ \hline 
$\sqrt{k}v^{1,0}$ &     & $0$ & $0$ & $0$ & $0$ & $0$ & $0$ 
\\ \hline 
$\sqrt{k}v^{1,1}$ &     & $-0.0166421$ & $-0.0234621$ & $-0.0269938$ & 
$-0.0291172$ & $-0.030423$ & $-0.0312168$ 
\\ \hline 
$\sqrt{k}v^{1,2}$ &     & $0$ & $0$ & $0$ & $0$ & $0$ & $0$ 
\\ \hline 
$\sqrt{k}v^{2,1}$ &     &     &     &     &     &     &     
\\ \hline 
$\sqrt{k}v^{2,2}$ &     &     &     &     &     &     &    
\\ \hline 
$\sqrt{k}v^{2,3}$ &     &     &     &     &     &     &     
\\ \hline 
$kv^{0,1,0}$ &     & $0.00174226$ & $-0.00215824$ & 
$-0.00455361$ & $-0.00627619$ & $-0.00757253$ & $-0.00858088$ 
\\ \hline 
$kv^{0,1,1}$ &     & $0$ & $0$ & $0$ & $0$ & $0$ & $0$ 
\\ \hline 
$kv^{0,1,2}$ &     & $0.0119636$ & $0.00937116$ & $0.00786394$ & 
$0.00678458$ & $0.00597277$ & $0.00533992$ 
\\ \hline 
$kv^{1,0,1}$ &     & $-0.0136868$ & $-0.0208768$ & 
$-0.0261862$ & $-0.030628$ & $-0.0344043$ & $-0.0376524$ 
\\ \hline 
$kv^{1,1,0}$ &     & $0$ & $0$ & $0$ & $0$ & $0$ & $0$ 
\\ \hline 
$kv^{1,1,1}$ &     & $0.0116287$ & $0.0186019$ & 
$0.0239239$ & $0.0283927$ & $0.0321994$ & $0.0354798$ 
\\ \hline 
$kv^{1,1,2}$ &     & $0$ & $0$ & $0$ & $0$ & $0$ & $0$ 
\\ \hline 
$kv^{1,2,1}$ &     & $0.00785285$ & $0.0123799$ & $0.0158031$ & 
$0.0186746$ & $0.0211193$ & $0.023225$ 
\\ \hline 
$kv^{1,2,2}$ &     & $0$ & $0$ & $0$ & $0$ & $0$ & $0$ 
\\ \hline 
$kv^{1,2,3}$ &     & $0.00730255$ & $0.00831474$ & 
$0.00805819$ & $0.00756586$ & $0.00704831$ & $0.00656395$ 
\\ \hline 
$kv^{2,1,0}$ &     &     &     &     &     &     &     \\ \hline 
$kv^{2,1,1}$ &     &     &     &     &     &     &     \\ \hline 
$kv^{2,1,2}$ &     &     &     &     &     &     &     \\ \hline 
$kv^{2,2,1}$ &     &     &     &     &     &     &     \\ \hline 
$kv^{2,2,2}$ &     &     &     &     &     &     &     \\ \hline 
$kv^{2,2,3}$ &     &     &     &     &     &     &     \\ \hline 
$kv^{2,3,2}$ &     &     &     &     &     &     &     \\ \hline 
$kv^{2,3,3}$ &     &     &     &     &     &     &     \\ \hline 
$kv^{2,3,4}$ &     &     &     &     &     &     &     \\ \hline 
$\beta^0$ &     & $-0.0399187$ & $-0.0468744$ & $-0.0508617$ & 
$-0.0534163$ & $-0.0551511$ & $-0.0563783$ 
\\ \hline 
$\beta^1$ &     & $-0.0428108$ & $-0.0593731$ & $-0.0688178$ & 
$-0.0756665$ & $-0.0809683$ & $-0.0852303$ 
\\ \hline 
$\beta^2$ &     &     &     &     &     &     &     
\\ \hline 
$w^0$ &     & $0.0152634$ & $0.0170702$ & $0.0180734$ & $0.0186733$ & $0.0190507$ & $0.0192958$ 
\\ \hline
$w^1$ &     & $0.0147015$ & $0.0199323$ & $0.0227881$ & $0.0248318$ & 
$0.0264046$ & $0.0276651$ 
\\ \hline 
$w^2$ &     &     &     &     &     &     &     
\\ \hline
$R^{[5/2,5]}_2$ &     & $0.697985$ & $0.787555$ & $0.809146$ & $0.822861$ & 
$0.833042$ & $0.841123$
\\ \hline
\end{tabular}
}

\begin{table}[htbp]
\label{tab2-2-2}
{\tiny
\begin{tabular}{|c|c|c|c|c|c|c|c|c|c|c|c|c|c|c|c|c|c|c|}\hline
 & $k=12$ & $k=14$ & $k=17$ &$k=50$ & $k=1000$ & $k=100000$ & $k=100000000$ 
\\ \hline
$t^0$ & $0.169426$ & $0.172746$ & $0.175698$ & $0.180675$ & 
$0.180581$ & $0.180531$ & $0.180530$ 
\\ \hline
$t^1$ & $0.361366$ & $0.367315$ & $0.372586$ & $0.382136$ & 
$0.382989$ & $0.382963$ & $0.382962$ 
\\ \hline 
$t^2$ & $0.295594$ & $0.293978$ & $0.292218$ & $0.287398$ & 
$0.285544$ & $0.285444$ & $0.285443$ 
\\ \hline 
$\sqrt{k}u^{0,1}$ & $0.000220801$ & 
$0.000176212$ & $0.000128004$ & $0.0000149582$ & 
$1.19311\times 10^{-8}$ & $1.12789\times 10^{-13}$ & 
$-1.44719\times 10^{-8}$ 
\\ \hline 
$\sqrt{k}u^{1,0}$ & $-0.100828$ & $-0.0967899$ & $-0.0914015$ & 
$-0.0606565$ & $-0.0144187$ & $-0.00144609$ & $-0.0000457406$ 
\\ \hline 
$\sqrt{k}u^{1,1}$ & $0$ & 
$0$ & $0$ & $0$ & $0$ & $0$ & $0$ \\ \hline 
$\sqrt{k}u^{1,2}$ & $-0.0910103$ & $-0.0844522$ & $-0.0769076$ & 
$-0.0455348$ & $-0.0102262$ & $-0.00102257$ & $-0.0000323221$ 
\\ \hline 
$\sqrt{k}u^{2,1}$ & $-0.0174166$ & $-0.0180781$ & 
$-0.0183917$ & $-0.0146409$ & $-0.00369371$ & 
$-0.000371329$ & $-0.0000117476$ 
\\ \hline 
$\sqrt{k}u^{2,2}$ & $0$ & 
$0$ & $0$ & $0$ & $0$ & $0$ & $0$ 
\\ \hline 
$\sqrt{k}u^{2,3}$ & $-0.0216297$ & $-0.0211472$ & 
$-0.0201924$ & $-0.0132189$ & $-0.00303102$ & 
$-0.000303204$ & $-9.58324\times 10^{-6}$ 
\\ \hline 
$\sqrt{k}v^{0,1}$ & $0$ & 
$0$ & $0$ & $0$ & $0$ & $0$ & $0$ 
\\ \hline 
$\sqrt{k}v^{1,0}$ & $0$ & 
$0$ & $0$ & $0$ & $0$ & $0$ & $0$ 
\\ \hline 
$\sqrt{k}v^{1,1}$ & $-0.0289082$ & $-0.0288156$ & 
$-0.0282087$ & $-0.0204471$ & $-0.00501921$ & 
$-0.000504157$ & $-0.0000159397$ 
\\ \hline 
$\sqrt{k}v^{1,2}$ & $0$ & 
$0$ & $0$ & $0$ & $0$ & $0$ & $0$ 
\\ \hline 
$\sqrt{k}v^{2,1}$ & $0$ & 
$0$ & $0$ & $0$ & $0$ & $0$ & $0$ 
\\ \hline 
$\sqrt{k}v^{2,2}$ & $-0.024225$ & 
$-0.0251935$ & $-0.0259052$ & $-0.0228624$ & 
$-0.00642731$ & $-0.000650811$ & $-0.0000205863$ 
\\ \hline 
$\sqrt{k}v^{2,3}$ & $0$ & 
$0$ & $0$ & $0$ & $0$ & $0$ & $0$ 
\\ \hline 
$kv^{0,1,0}$ & $-0.0093589$ & $-0.0103545$ & 
$-0.0113574$ & $-0.0140093$ & 
$-0.0149428$ & $-0.0149815$ & $-0.0149819$ 
\\ \hline 
$kv^{0,1,1}$ & $0$ & 
$0$ & $0$ & $0$ & $0$ & $0$ & $0$ 
\\ \hline 
$kv^{0,1,2}$ & $0.00466621$ & $0.00396889$ & 
$0.00324558$ & $0.00108347$ & $0.0000532027$ & 
$5.31205\times 10^{-7}$ & $5.31116\times 10^{-10}$ 
\\ \hline 
$kv^{1,0,1}$ & $-0.0414168$ & 
$-0.0451461$ & $-0.0494101$ & $-0.0648339$ & 
$-0.0736724$ & $-0.0741517$ & $-0.0741255$ 
\\ \hline 
$kv^{1,1,0}$ & $0$ & 
$0$ & $0$ & $0$ & $0$ & $0$ & $0$ 
\\ \hline 
$kv^{1,1,1}$ & $0.0389862$ & $0.0428163$ & 
$0.0472180$ & $0.0633867$ & 
$0.0728707$ & $0.0733907$ & $0.0733690$ 
\\ \hline 
$kv^{1,1,2}$ & $0$ & 
$0$ & $0$ & $0$ & $0$ & $0$ & $0$ 
\\ \hline 
$kv^{1,2,1}$ & $0.025528$ & $0.0279752$ & $0.0307837$ & 
$0.0410573$ & $0.0470461$ & $0.0473736$ & $0.0473422$ 
\\ \hline 
$kv^{1,2,2}$ & $0$ & 
$0$ & $0$ & $0$ & $0$ & $0$ & $0$ 
\\ \hline 
$kv^{1,2,3}$ & $0.0053121$ & $0.0046877$ & 
$0.0039728$ & $0.00145488$ & 
$0.0000742872$ & $7.43245\times 10^{-7}$ & $7.42587\times 10^{-10}$ 
\\ \hline 
$kv^{2,1,0}$ & $-0.00459930$ & $-0.00424815$ & 
$-0.00380153$ & $-0.00172786$ & 
$-0.000101270$ & $-1.02111\times 10^{-6}$ & $-1.02082\times 10^{-9}$ 
\\ \hline 
$kv^{2,1,1}$ & $0$ & 
$0$ & $0$ & $0$ & $0$ & $0$ & $0$ 
\\ \hline 
$kv^{2,1,2}$ & $-0.0334439$ & $-0.0390379$ & 
$-0.0459434$ & $-0.0779293$ & 
$-0.104242$ & $-0.105921$ & $-0.105969$ 
\\ \hline 
$kv^{2,2,1}$ & $0$ & 
$0$ & $0$ & $0$ & $0$ & $0$ & $0$ 
\\ \hline 
$kv^{2,2,2}$ & $0.0107035$ & $0.0140437$ & 
$0.0181942$ & $0.0376032$ & $0.0536692$ & $0.0546957$ & 
$0.0547196$ 
\\ \hline 
$kv^{2,2,3}$ & $0$ & 
$0$ & $0$ & $0$ & $0$ & $0$ & $0$ 
\\ \hline 
$kv^{2,3,2}$ & $0.0211825$ & $0.0257547$ & $0.0314171$ & 
$0.0577668$ & $0.0795089$ & $0.0808974$ & $0.0809434$ 
\\ \hline 
$kv^{2,3,3}$ & $0$ & $0$ & $0$ & $0$ & $0$ & $0$ & $0$ 
\\ \hline 
$kv^{2,3,4}$ & $0.00118433$ & $0.00113442$ & 
$0.00104673$ & $0.00049862$ & 
$0.0000293805$ & $2.96340\times 10^{-7}$ & $2.96127\times 10^{-10}$ 
\\ \hline 
$\beta^0$ & $-0.0575724$ & $-0.0583195$ & $-0.0588883$ & 
$-0.0589482$ & $-0.0578371$ & 
$-0.0577554$ & $-0.0577546$ 
\\ \hline 
$\beta^1$ & 
$-0.090402$ & $-0.0947988$ & 
$-0.0995262$ & $-0.114572$ & $-0.122119$ & 
$-0.122512$ & $-0.122519$ 
\\ \hline 
$\beta^2$ & 
$-0.0379856$ & $-0.0429174$ & $-0.0487671$ & 
$-0.072916$ & $-0.0902554$ & 
$-0.0913073$ & $-0.0913180$ \\ \hline 
$w^0$ & 
$0.0191403$ & $0.0191689$ & 
$0.0191142$ & $0.0182665$ & $0.0173586$ & 
$0.0173002$ & $0.0172996$ 
\\ \hline 
$w^1$ & 
$0.0285543$ & $0.029739$ & 
$0.0309979$ & $0.0348483$ & $0.0366107$ & 
$0.0366971$ & $0.0366979$ 
\\ \hline 
$w^2$ & 
$0.0103462$ & $0.0119149$ & 
$0.0137721$ & $0.0214459$ & $0.0270102$ & 
$0.0273496$ & $0.0273530$ 
\\ \hline
$R^{[5/2,5]}_2$ & $0.861307$ & $0.870218$ & $0.88037$ & $0.919601$ & 
$0.946874$ & $0.948536$ & $0.948553$
\\ \hline
\end{tabular}
}
\caption{${\cal T}^{[5/2,5]}_2$ and $R^{[5/2,5]}_2$ ($J=1$)}
\end{table}

\begin{table}[htbp]
\label{tabn1}
{\tiny
\begin{tabular}{|c|c|c|c|c|c|c|c|c|}\hline
 & & $k=2$ & $k=3$ & $k=4$ & $k=5$ & $k=6$ & $k=7$ & $k=8$ 
\\ \hline
$\wt{t}^0$ & linear &
$0$ & $0$ & $0$ & $0$ & $0$ & $0$ & $0$
\\ \cline{2-9}
 & quadratic &
$0$ & $0$ & $0$ & $0$ & $0$ & $0$ & $0$
\\ \hline
$\wt{t}^1$ & linear &
$0$ & $0$ & $0$ & $0$ & $0$ & $0$ & $0$
\\ \cline{2-9}
 & quadratic &
$0$ & $0$ & $0$ & $0$ & $0$ & $0$ & $0$
\\ \hline
$\wt{u}^0$ & linear &
$-0.115553$ & $-0.096259$ & $-0.0996669$ & $-0.0999901$ & 
$-0.0994142$ & $-0.0985275$ & $-0.0975507$
\\ \cline{2-9}
 & quadratic &
$0.0801271$ & $0.0760034$ & $0.0774179$ & $0.0773437$ & 
$0.0767866$ & $0.0760797$ & $0.075348$
\\ \cline{2-9}
 & sum & $-0.0354259$ & $-0.0202556$ & $-0.022249$ & $-0.0226464$ & 
$-0.0226276$ & $-0.0224478$ & $-0.0222027$
\\ \cline{2-9}
 & sum/linear & $0.306577$ & $0.210428$ & $0.223234$ & $0.226486$ & 
$0.227609$ & $0.227833$ & $0.227602$
\\ \hline
$\wt{u}^1$ & linear &
    & $-0.0449295$ & $-0.0409692$ & $-0.040177$ & 
$-0.0399862$ & $-0.0399995$ & $-0.040098$
\\ \cline{2-9}
 & quadratic &
    & $0.0328215$ & $0.0327684$ & $0.032912$ & $0.0330957$ & 
$0.0332766$ & $0.0334437$
\\ \cline{2-9}
 & sum &     & $-0.012108$ & $-0.0082008$ & $-0.007265$ & $-0.0068905$ & 
$-0.0067229$ & $-0.0066543$
\\ \cline{2-9}
 & sum/linear &     & $0.269489$ & $0.20017$ & $0.180825$ & 
$0.172322$ & $0.168075$ & $0.165951$
\\ \hline
$\sqrt{k}\wt{u}^{0,0}$ & linear &
$0$ & $0$ & $0$ & $0$ & $0$ & $0$ & $0$
\\ \cline{2-9}
 & quadratic &
$0$ & $0$ & $0$ & $0$ & $0$ & $0$ & $0$
\\ \hline
$\sqrt{k}\wt{u}^{1,0}$ & linear &
    & $0$ & $0$ & $0$ & $0$ & $0$ & $0$
\\ \cline{2-9}
 & quadratic &
    & $0$ & $0$ & $0$ & $0$ & $0$ & $0$
\\ \hline
$\sqrt{k}\wt{u}^{1,1}$ & linear &
    & $0.00395757$ & $0.00316496$ & $0.00326707$ & 
$0.00337508$ & $0.00344044$ & $0.00347031$
\\ \cline{2-9}
 & quadratic &
    & $-0.00223712$ & $-0.00176369$ & $-0.00143251$ & $-0.0011936$ & 
$-0.00101502$ & $-0.000877389$
\\ \hline
$\sqrt{k}\wt{u}^{1,2}$ & linear &
    & $0$ & $0$ & $0$ & $0$ & $0$ & $0$
\\ \cline{2-9}
 & quadratic &
    & $0$ & $0$ & $0$ & $0$ & $0$ & $0$
\\ \hline
\end{tabular}
}

{\tiny
\begin{tabular}{|c|c|c|c|c|c|c|c|}\hline
 & & $k=9$ & $k=10$ & $k=50$ & $k=1000$ & $k=100000$ & $k=100000000$ 
\\ \hline
$\wt{t}^0$ & linear &
$0$ & $0$ & $0$ & $0$ & $0$ & $0$ 
\\ \cline{2-8}
 & quadratic &
$0$ & $0$ & $0$ & $0$ & $0$ & $0$
\\ \hline
$\wt{t}^1$ & linear &
$0$ & $0$ & $0$ & $0$ & $0$ & $0$
\\ \cline{2-8}
 & quadratic &
$0$ & $0$ & $0$ & $0$ & $0$ & $0$
\\ \hline
$\wt{u}^0$ & linear &
$-0.0965762$ & $-0.0956433$ & $-0.0830391$ & 
$-0.0777582$ & $-0.0774489$ & $-0.0774462$ 
\\ \cline{2-8}
 & quadratic &
$0.0746406$ & $0.0739763$ & 
$0.065545$ & $0.0621793$ & $0.0619843$ & $0.0619821$ 
\\ \cline{2-8}
 & sum & $-0.0219356$ & $-0.021667$ & $-0.0174941$ & $-0.0155789$ & 
$-0.0154646$ & $-0.0154641$
\\ \cline{2-8}
 & sum/linear & $0.227133$ & $0.22654$ & $0.210673$ & $0.200351$ & 
$0.199675$ & $0.199675$
\\ \hline
$\wt{u}^1$ & linear &
$-0.0402346$ & $-0.040387$ & 
$-0.0432225$ & $-0.0446278$ & $-0.0447131$ & $-0.0447136$ 
\\ \cline{2-8}
 & quadratic &
$0.0335948$ & $0.0337304$ & $0.0352303$ & 
$0.035756$ & $0.0357852$ & $0.0357854$
\\ \cline{2-8}
 & sum & $-0.0066398$ & $-0.0066566$ & $-0.0079922$ & $-0.0088718$ & 
$-0.0089279$ & $-0.0089282$
\\ \cline{2-8}
 & sum/linear & $0.165027$ & $0.16482$ & $0.184908$ & $0.198795$ & 
$0.199671$ & $0.199675$
\\ \hline
$\sqrt{k}\wt{u}^{0,0}$ & linear &
$0$ & $0$ & $0$ & $0$ & $0$ & $0$ 
\\ \cline{2-8}
 & quadratic &
$0$ & $0$ & $0$ & $0$ & $0$ & $0$ 
\\ \hline
$\sqrt{k}\wt{u}^{1,0}$ & linear &
$0$ & $0$ & $0$ & $0$ & $0$ & $0$ 
\\ \cline{2-8}
 & quadratic &
$0$ & $0$ & $0$ & $0$ & $0$ & $0$ 
\\ \hline
$\sqrt{k}\wt{u}^{1,1}$ & linear &
$0.0034751$ & $0.0001095$ & $0.00222367$ & $0.000537711$ & 
$0.000053975$ & $1.70688\times 10^{-6}$ 
\\ \cline{2-8}
 & quadratic &
$-0.0007686$ & $-0.00068081$ & 
$-0.0000830692$ & $-1.01643\times 10^{-6}$ & $-9.95555\times 10^{-10}$ & 
$-3.22551\times 10^{-14}$
\\ \hline
$\sqrt{k}\wt{u}^{1,2}$ & linear &
$0$ & $0$ & $0$ & $0$ & $0$ & $0$ 
\\ \cline{2-8}
 & quadratic &
$0$ & $0$ & $0$ & $0$ & $0$ & $0$ 
\\ \hline
\end{tabular}
}
\caption{The values of the equations of motion of 
$\wt{t}, \wt{u}$ and $\wt{u}^a$ for ${\cal T}^{[5/2,5]}_1$ ($J=1/2$)}
\end{table}

\begin{table}[htbp]
\label{tabn2-1}
{\tiny
\begin{tabular}{|c|c|c|c|c|c|c|c|c|}\hline
 & & $k=4$ & $k=5$ & $k=6$ & $k=7$ & $k=8$ & $k=9$ & $k=10$ 
\\ \hline
$\wt{t}^0$ & linear &
$0$ & $0$ & $0$ & $0$ & $0$ & $0$ & $0$ 
\\ \cline{2-9}
 & quadratic & 
$0$ & $0$ & $0$ & $0$ & $0$ & $0$ & $0$ 
\\ \hline
$\wt{t}^1$ & linear & 
$0$ & $0$ & $0$ & $0$ & $0$ & $0$ & $0$ 
\\ \cline{2-9}
 & quadratic & 
$0$ & $0$ & $0$ & $0$ & $0$ & $0$ & $0$ 
\\ \hline
$\wt{t}^2$ & linear &
$0$ & $0$ & $0$ & $0$ & $0$ & $0$ & $0$ 
\\ \cline{2-9}
 & quadratic & 
$0$ & $0$ & $0$ & $0$ & $0$ & $0$ & $0$ 
\\ \hline
$\wt{u}^0$ & linear &
 $-0.122452$ & $-0.126636$ & $-0.127711$ & $-0.128512$ & 
$-0.128968$ & $-0.129225$ & $-0.129358$ 
\\ \cline{2-9}
 & quadratic & 
$0.0953565$ & $0.0973952$ & $0.097417$ & $0.0974845$ & $0.0974351$ & 
$0.0973243$ & $0.0971819$ 
\\ \cline{2-9}
 & sum & $-0.0270955$ & $-0.0292408$ & $-0.030294$ & $-0.0310275$ & 
$-0.0315329$ & $-0.0319007$ & $-0.0321761$
\\ \cline{2-9}
 & sum/linear & $0.221274$ & $0.230904$ & $0.237207$ & $0.241437$ & 
$0.244502$ & $0.246862$ & $0.248737$
\\ \hline
$\wt{u}^1$ & linear & 
$-0.0343375$ & $-0.0297219$ & $-0.0277301$ & 
$-0.0263031$ & $-0.0252966$ & $-0.0245495$ & 
$-0.0239751$ 
\\ \cline{2-9}
 & quadratic & 
$0.0201982$ & $0.0194408$ & $0.0178695$ & $0.0173807$ & $0.0169976$ & 
$0.0221006$ & $0.0219371$ 
\\ \cline{2-9}
 & sum & $-0.0141393$ & $-0.0102811$ & $-0.0098606$ & $-0.0089224$ & 
$-0.008299$ & $-0.0024489$ & $-0.002038$
\\ \cline{2-9}
 & sum/linear & $0.411774$ & $0.34591$ & $0.355592$ & $0.339215$ & 
$0.328068$ & $0.0997536$ & $0.0850049$
\\ \hline
$\wt{u}^2$ & linear &
    &     &     &     &     &     &     
\\ \cline{2-9}
 & quadratic & 
    &     &     &     &     &     &     
\\ \hline
$\sqrt{k}\wt{u}^{0,1}$ & linear &
$0$ & $0$ & $0$ & $0$ & $0$ & $0$ & $0$ 
\\ \cline{2-9}
 & quadratic & 
$0$ & $0$ & $0$ & $0$ & $0$ & $0$ & $0$ 
\\ \hline
$\sqrt{k}\wt{u}^{1,0}$ & linear & 
$0$ & $0$ & $0$ & $0$ & $0$ & $0$ & $0$ 
\\ \cline{2-9}
 & quadratic & 
$0$ & $0$ & $0$ & $0$ & $0$ & $0$ & $0$ 
\\ \hline
$\sqrt{k}\wt{u}^{1,1}$ & linear &
$-0.00108766$ & $-0.0013597$ & $-0.00123209$ & 
$-0.000999576$ & $-0.000786557$ & $-0.00060636$ & $-0.000457082$
\\ \cline{2-9}
 & quadratic & 
$-0.0021537$ & $-0.00173246$ & $-0.00143638$ & $-0.00122214$ & 
$-0.00106054$ & $-0.00008257$ & $-0.000011728$ 
\\ \hline
$\sqrt{k}\wt{u}^{1,2}$ & linear & 
$0$ & $0$ & $0$ & $0$ & $0$ & $0$ & $0$ 
\\ \cline{2-9}
 & quadratic & 
$0$ & $0$ & $0$ & $0$ & $0$ & $0$ & $0$ 
\\ \hline
$\sqrt{k}\wt{u}^{2,1}$ & linear &
    &     &     &     &     &     &     
\\ \cline{2-9}
 & quadratic & 
    &     &     &     &     &     &     
\\ \hline
$\sqrt{k}\wt{u}^{2,2}$ & linear &
    &     &     &     &     &     &     
\\ \cline{2-9}
 & quadratic & 
    &     &     &     &     &     &     
\\ \hline
$\sqrt{k}\wt{u}^{2,3}$ & linear &
    &     &     &     &     &     &     
\\ \cline{2-9}
 & quadratic & 
    &     &     &     &     &     &     
\\ \hline
\end{tabular}
}

{\tiny
\begin{tabular}{|c|c|c|c|c|c|c|c|c|}\hline
 & & $k=12$ & $k=14$ & $k=17$ &$k=50$ & 
$k=1000$ & $k=100000$ & $k=100000000$ 
\\ \hline
$\wt{t}^0$ & linear &
$0$ & $0$ & $0$ & $0$ & $0$ & $0$ & $0$ 
\\ \cline{2-9}
 & quadratic & 
$0$ & $0$ & $0$ & $0$ & $0$ & $0$ & $0$ 
\\ \hline
$\wt{t}^1$ & linear & 
$0$ & $0$ & $0$ & $0$ & $0$ & $0$ & $0$ 
\\ \cline{2-9}
 & quadratic & 
$0$ & $0$ & $0$ & $0$ & $0$ & $0$ & $0$ 
\\ \hline
$\wt{t}^2$ & linear &
$0$ & $0$ & $0$ & $0$ & $0$ & $0$ & $0$ 
\\ \cline{2-9}
 & quadratic & 
$0$ & $0$ & $0$ & $0$ & $0$ & $0$ & $0$ 
\\ \hline
$\wt{u}^0$ & linear &
$-0.124279$ & $-0.119976$ & $-0.118333$ & $-0.110253$ & 
$-0.103677$ & $-0.103266$ & $-0.103262$ 
\\ \cline{2-9}
 & quadratic & 
$0.0939498$ & $0.0936305$ & $0.0925177$ & $0.0871328$ & 
$0.0829048$ & $0.0826455$ & $0.0826428$
\\ \cline{2-9}
 & sum & $-0.0303292$ & $-0.0263455$ & $-0.0258153$ & 
$-0.0231202$ & $-0.0207722$ & $-0.0206205$ & $-0.0206192$
\\ \cline{2-9}
 & sum/linear & $0.244041$ & $0.21959$ & $0.218158$ & 
$0.209701$ & $0.200355$ & $0.199683$ & $0.199678$
\\ \hline
$\wt{u}^1$ & linear & 
$-0.0307889$ & $-0.0300733$ & $-0.0304782$ & $-0.033354$ & 
$-0.0363096$ & $-0.0365065$ & $-0.0365085$ 
\\ \cline{2-9}
 & quadratic & 
$0.0243392$ & $0.0248842$ & $0.0252412$ & $0.0272825$ & 
$0.0291017$ & $0.0292175$ & $0.0292186$
\\ \cline{2-9}
 & sum & $-0.0064497$ & $-0.0051891$ & $-0.005237$ & 
$-0.0060715$ & $-0.0072079$ & $-0.007289$ & $-0.0072899$
\\ \cline{2-9}
 & sum/linear & $0.209481$ & $0.172548$ & $0.171828$ & 
$0.182032$ & $0.198512$ & $0.199663$ & $0.199677$
\\ \hline
$\wt{u}^2$ & linear &
$0.0119620$ & $0.0127145$ & $0.0133954$ & $0.0153322$ & 
$0.0162758$ & $0.0163266$ & $0.0163271$ 
\\ \cline{2-9}
 & quadratic & 
$-0.0122473$ & $-0.0129385$ & $-0.011892$ & $-0.0130049$ & 
$-0.0130764$ & $-0.0130671$ & $-0.013067$
\\ \cline{2-9}
 & sum & $-0.0002853$ & $-0.000224$ & $0.0015034$ & $0.0023273$ & 
$0.0031994$ & $0.0032595$ & $0.0032601$
\\ \cline{2-9}
 & sum/linear & $-0.0238505$ & $-0.0176177$ & $0.112233$ & 
$0.151792$ & $0.196574$ & $0.199644$ & $0.199674$
\\ \hline
$\sqrt{k}\wt{u}^{0,1}$ & linear &
$0$ & $0$ & $0$ & $0$ & $0$ & $0$ & $0$ 
\\ \cline{2-9}
 & quadratic & 
$0$ & $0$ & $0$ & $0$ & $0$ & $0$ & $0$ 
\\ \hline
$\sqrt{k}\wt{u}^{1,0}$ & linear & 
$0$ & $0$ & $0$ & $0$ & $0$ & $0$ & $0$ 
\\ \cline{2-9}
 & quadratic & 
$0$ & $0$ & $0$ & $0$ & $0$ & $0$ & $0$ 
\\ \hline
$\sqrt{k}\wt{u}^{1,1}$ & linear &
$0.00236956$ & $0.00144474$ & $0.00159835$ & $0.0015727$ & 
$0.000436163$ & $0.000044067$ & $1.39366\times 10^{-6}$ 
\\ \cline{2-9}
 & quadratic & 
$-0.000292312$ & $-0.000235668$ & $-0.000180359$ & $-0.0000377101$ & 
$-4.22556\times 10^{-7}$ & $-4.22259\times 10^{-10}$ & 
$-1.33525\times 10^{-14}$
\\ \hline
$\sqrt{k}\wt{u}^{1,2}$ & linear & 
$0$ & $0$ & $0$ & $0$ & $0$ & $0$ & $0$
\\ \cline{2-9}
 & quadratic & 
$0$ & $0$ & $0$ & $0$ & $0$ & $0$ & $0$
\\ \hline
$\sqrt{k}\wt{u}^{2,1}$ & linear &
$0$ & $0$ & $0$ & $0$ & $0$ & $0$ & $0$
\\ \cline{2-9}
 & quadratic & 
$0$ & $0$ & $0$ & $0$ & $0$ & $0$ & $0$
\\ \hline
$\sqrt{k}\wt{u}^{2,2}$ & linear &
$-0.00234631$ & $-0.00180218$ & 
$-0.00178822$ & $-0.00135708$ & $-0.000339651$ & 
$-0.0000341358$ & $-1.07953\times 10^{-6}$ 
\\ \cline{2-9}
 & quadratic & 
$0.000918385$ & $0.000782397$ & $0.000634075$ & $0.000168119$ & 
$2.23462\times 10^{-6}$ & $2.25649\times 10^{-9}$ & 
$7.13634\times 10^{-14}$
\\ \hline
$\sqrt{k}\wt{u}^{2,3}$ & linear &
$0$ & $0$ & $0$ & $0$ & $0$ & $0$ & $0$
\\ \cline{2-9}
 & quadratic & 
$0$ & $0$ & $0$ & $0$ & $0$ & $0$ & $0$
\\ \hline
\end{tabular}
}
\caption{The values of the equations of motion of 
$\wt{t}, \wt{u}$ and $\wt{u}^a$ for ${\cal T}^{[5/2,5]}_1$ ($J=1$)}
\end{table}

\begin{table}[htbp]
\label{tabn2-2}
{\tiny
\begin{tabular}{|c|c|c|c|c|c|c|c|c|}\hline
 & & $k=4$ & $k=5$ & $k=6$ & $k=7$ & $k=8$ & $k=9$ & $k=10$ 
\\ \hline
$\wt{t}^0$ & linear &
    & $0$ & $0$ & $0$ & $0$ & $0$ & $0$
\\ \cline{2-9}
 & quadratic & 
    & $0$ & $0$ & $0$ & $0$ & $0$ & $0$
\\ \hline
$\wt{t}^1$ & linear &
    & $0$ & $0$ & $0$ & $0$ & $0$ & $0$
\\ \cline{2-9}
 & quadratic & 
    & $0$ & $0$ & $0$ & $0$ & $0$ & $0$
\\ \hline
$\wt{t}^2$ & linear &
    & $0$ & $0$ & $0$ & $0$ & $0$ & $0$
\\ \cline{2-9}
 & quadratic & 
    & $0$ & $0$ & $0$ & $0$ & $0$ & $0$
\\ \hline
$\wt{u}^0$ & linear &
    & $-0.0592968$ & $-0.0658236$ & $-0.0691706$ & 
$-0.0709667$ & $-0.0719414$ & $-0.0724533$
\\ \cline{2-9}
 & quadratic & 
    & $0.0479373$ & $0.0522838$ & $0.0544314$ & $0.0555162$ & 
$0.0560578$ & $0.0563016$
\\ \cline{2-9}
 & sum &     & $-0.0113595$ & $-0.0135398$ & $-0.0147392$ & 
$-0.0154505$ & $-0.0158836$ & $-0.0161517$
\\ \cline{2-9}
 & sum/linear &     & $0.19157$ & $0.205698$ & $0.213085$ & 
$0.217715$ & $0.220785$ & $0.222926$
\\ \hline
$\wt{u}^1$ & linear &
    & $-0.0269091$ & $-0.03365$ & $-0.0363017$ & 
$-0.0378647$ & $-0.03892$ & $-0.0396928$
\\ \cline{2-9}
 & quadratic & 
    & $0.0277842$ & $0.0337662$ & $0.0361454$ & $0.0375423$ & 
$0.0317393$ & $0.0322759$
\\ \cline{2-9}
 & sum &     & $0.0008751$ & $0.0001162$ & $-0.0001563$ & 
$-0.0003224$ & $-0.0071807$ & $-0.0074169$
\\ \cline{2-9}
 & sum/linear &     & $-0.0325206$ & $-0.00345319$ & $0.00430558$ & 
$0.00851453$ & $0.184499$ & $0.186858$
\\ \hline
$\wt{u}^2$ & linear &
    &     &     &     &     &     &    
\\ \cline{2-9}
 & quadratic & 
    &     &     &     &     &     &    
\\ \hline
$\sqrt{k}\wt{u}^{0,1}$ & linear &
    & $0$ & $0$ & $0$ & $0$ & $0$ & $0$
\\ \cline{2-9}
 & quadratic & 
    & $0$ & $0$ & $0$ & $0$ & $0$ & $0$
\\ \hline
$\sqrt{k}\wt{u}^{1,0}$ & linear &
    & $0$ & $0$ & $0$ & $0$ & $0$ & $0$
\\ \cline{2-9}
 & quadratic & 
    & $0$ & $0$ & $0$ & $0$ & $0$ & $0$
\\ \hline
$\sqrt{k}\wt{u}^{1,1}$ & linear &
    & $0.000771072$ & $0.00222674$ & 
$0.00311876$ & $0.00363842$ & $0.00395313$ & $0.0041485$
\\ \cline{2-9}
 & quadratic & 
    & $-0.000322998$ & $-0.000396918$ & $-0.000395812$ & 
$-0.000374473$ & $-0.00127216$ & $-0.00122043$
\\ \hline
$\sqrt{k}\wt{u}^{1,2}$ & linear &
    & $0$ & $0$ & $0$ & $0$ & $0$ & $0$
\\ \cline{2-9}
 & quadratic & 
    & $0$ & $0$ & $0$ & $0$ & $0$ & $0$
\\ \hline
$\sqrt{k}\wt{u}^{2,1}$ & linear &
    &     &     &     &     &     &    
\\ \cline{2-9}
 & quadratic & 
    &     &     &     &     &     &    
\\ \hline
$\sqrt{k}\wt{u}^{2,2}$ & linear &
    &     &     &     &     &     &    
\\ \cline{2-9}
 & quadratic & 
    &     &     &     &     &     &    
\\ \hline
$\sqrt{k}\wt{u}^{2,3}$ & linear &
    &     &     &     &     &     &    
\\ \cline{2-9}
 & quadratic & 
    &     &     &     &     &     &    
\\ \hline
\end{tabular}
}

{\tiny
\begin{tabular}{|c|c|c|c|c|c|c|c|c|}\hline
 & & $k=12$ & $k=14$ & $k=17$ &$k=50$ & 
$k=1000$ & $k=100000$ & $k=100000000$ 
\\ \hline
$\wt{t}^0$ & linear &
$0$ & $0$ & $0$ & $0$ & $0$ & $0$ & $0$ 
\\ \cline{2-9}
 & quadratic & 
$0$ & $0$ & $0$ & $0$ & $0$ & $0$ & $0$ 
\\ \hline
$\wt{t}^1$ & linear &
$0$ & $0$ & $0$ & $0$ & $0$ & $0$ & $0$ 
\\ \cline{2-9}
 & quadratic & 
$0$ & $0$ & $0$ & $0$ & $0$ & $0$ & $0$ 
\\ \hline
$\wt{t}^2$ & linear &
$0$ & $0$ & $0$ & $0$ & $0$ & $0$ & $0$ 
\\ \cline{2-9}
 & quadratic & 
$0$ & $0$ & $0$ & $0$ & $0$ & $0$ & $0$ 
\\ \hline
$\wt{u}^0$ & linear &
$-0.0677078$ & $-0.0670125$ & $-0.0658376$ & $-0.0585395$ & 
$-0.0520464$ & $-0.051635$ & $-0.0516308$
\\ \cline{2-9}
 & quadratic & 
$0.0534401$ & $0.052787$ & $0.0517889$ & 
$0.0462182$ & $0.041609$ & $0.0413243$ & $0.0413214$
\\ \cline{2-9}
 & sum & $-0.0142677$ & $-0.0142255$ & $-0.0140487$ & 
$-0.0123213$ & $-0.0104374$ & $-0.0103107$ & $-0.0103094$
\\ \cline{2-9}
 & sum/linear & $0.210725$ & $0.212281$ & $0.213384$ & 
$0.210478$ & $0.20054$ & $0.199684$ & $0.199675$
\\ \hline
$\wt{u}^1$ & linear &
$-0.0368546$ & $-0.0370745$ & $-0.0372371$ & 
$-0.0371631$ & $-0.0365589$ & $-0.036509$ & $-0.0365085$
\\ \cline{2-9}
 & quadratic & 
$0.0305126$ & $0.0306218$ & $0.0306575$ & $0.0301119$ & 
$0.0292786$ & $0.0292193$ & $0.0292186$
\\ \cline{2-9}
 & sum & $-0.006342$ & $-0.0064527$ & $-0.0065796$ & 
$-0.0070512$ & $-0.0072803$ & $-0.0072897$ & $-0.0072899$
\\ \cline{2-9}
 & sum/linear & $0.172082$ & $0.174047$ & $0.176695$ & 
$0.189737$ & $0.199139$ & $0.199669$ & $0.199677$
\\ \hline
$\wt{u}^2$ & linear &
$-0.00922363$ & $-0.00984824$ & $-0.0105588$ & $-0.0135921$ & 
$-0.0161567$ & $-0.0163254$ & $-0.0163271$
\\ \cline{2-9}
 & quadratic & 
$0.0067855$ & $0.00717768$ & $0.0093856$ & $0.011643$ & 
$0.0129895$ & $0.0130662$ & $0.0130670$
\\ \cline{2-9}
 & sum & $-0.00243813$ & $-0.00267056$ & $-0.0011732$ & 
$-0.0019491$ & $-0.0031672$ & $-0.0032592$ & $-0.0032601$
\\ \cline{2-9}
 & sum/linear & $0.264335$ & $0.271171$ & $0.111111$ & 
$0.143399$ & $0.19603$ & $0.19964$ & $0.199674$
\\ \hline
$\sqrt{k}\wt{u}^{0,1}$ & linear &
$0$ & $0$ & $0$ & $0$ & $0$ & $0$ & $0$ 
\\ \cline{2-9}
 & quadratic & 
$0$ & $0$ & $0$ & $0$ & $0$ & $0$ & $0$ 
\\ \hline
$\sqrt{k}\wt{u}^{1,0}$ & linear &
$0$ & $0$ & $0$ & $0$ & $0$ & $0$ & $0$ 
\\ \cline{2-9}
 & quadratic & 
$0$ & $0$ & $0$ & $0$ & $0$ & $0$ & $0$ 
\\ \hline
$\sqrt{k}\wt{u}^{1,1}$ & linear &
$0.00249842$ & $0.00257273$ & 
$0.00257823$ & $0.00188571$ & $0.000440709$ & 
$0.0000440714$ & $1.39366\times 10^{-6}$
\\ \cline{2-9}
 & quadratic & 
$-0.000686175$ & $-0.000585444$ & $-0.000473995$ & $-0.000124947$ & 
$-1.66320\times 10^{-6}$ & $-1.67990\times 10^{-9}$ & 
$-5.31279\times 10^{-14}$\\ \hline
$\sqrt{k}\wt{u}^{1,2}$ & linear &
$0$ & $0$ & $0$ & $0$ & $0$ & $0$ & $0$
\\ \cline{2-9}
 & quadratic & 
$0$ & $0$ & $0$ & $0$ & $0$ & $0$ & $0$
\\ \hline
$\sqrt{k}\wt{u}^{2,1}$ & linear &
$0$ & $0$ & $0$ & $0$ & $0$ & $0$ & $0$
\\ \cline{2-9}
 & quadratic & 
$0$ & $0$ & $0$ & $0$ & $0$ & $0$ & $0$
\\ \hline
$\sqrt{k}\wt{u}^{2,2}$ & linear &
$0.00147773$ & $0.00148346$ & $0.00147774$ & $0.00122804$ & 
$0.000337650$ & $0.0000341339$ & $1.07953\times 10^{-6}$ 
\\ \cline{2-9}
 & quadratic & 
$-0.000511527$ & $-0.00043652$ & $-0.000354255$ & $-0.0000952753$ & 
$-1.28726\times 10^{-6}$ & $-1.30123\times 10^{-9}$ & 
$-4.11527\times 10^{-14}$
\\ \hline
$\sqrt{k}\wt{u}^{2,3}$ & linear &
$0$ & $0$ & $0$ & $0$ & $0$ & $0$ & $0$ 
\\ \cline{2-9}
 & quadratic & 
$0$ & $0$ & $0$ & $0$ & $0$ & $0$ & $0$ 
\\ \hline
\end{tabular}
}
\caption{The values of the equations of motion of 
$\wt{t}, \wt{u}$ and $\wt{u}^a$ for ${\cal T}^{[5/2,5]}_2$ ($J=1$)}
\end{table}
\end{center}

\clearpage
\newcommand{\J}[4]{{\sl #1} {\bf #2} (#3) #4}
\newcommand{\andJ}[3]{{\bf #1} (#2) #3}
\newcommand{\AP}{Ann.\ Phys.\ (N.Y.)}
\newcommand{\MPL}{Mod.\ Phys.\ Lett.}
\newcommand{\NP}{Nucl.\ Phys.}
\newcommand{\PL}{Phys.\ Lett.}
\newcommand{\PR}{Phys.\ Rev.}
\newcommand{\PRL}{Phys.\ Rev.\ Lett.}
\newcommand{\PTP}{Prog.\ Theor.\ Phys.}
\newcommand{\hepth}[1]{{\tt hep-th/#1}}


\begin{thebibliography}{99}

\bibitem{s1}
 A.\ Sen, ``{\it Descent relations among bosonic D-branes}'',
 \J{Int.\ J.\ Mod.\ Phys.}{A14}{1999}{4061}
 \hepth{9902105}

\bibitem{rscclmpt}
 A.\ Recknagel and V.\ Schomerus, ``{\it Boundary deformation theory
 and moduli spaces of D-branes}'',
 \hepth{9811237}, \J{\NP}{B545}{1999}{233};
 C.\ G.\ Callan, I.\ R.\ Klebanov, A.\ W.\ Ludwig and J.\ M.\ Maldacena, 
 ``{\it Exact solution of a boundary conformal field theory}'',
 \hepth{9402113}, \J{\NP}{B422}{1994}{417};
 J.\ Polchinski, and L.\ Thorlacius, 
 ``{\it Free fermion representation of a boundary conformal field
 theory}'',
 \hepth{9404008}, \J{\PR}{D50}{1994}{622}

\bibitem{w1}
 E.\ Witten, ``{\it Noncommutative geometry and string field theory}'',
 \J{\NP}{B268}{1986}{253}

\bibitem{boundary}
E.\ Witten,
``{\it On background independent open string field theory}'',
 \J{\PR}{D46}{5467}{1992}, \hepth{9208027} ;
E.\ Witten,
``{\it Some computations in background independent off-shell 
string theory}'',
 \J{\PR}{D47}{5467}{1992}, \hepth{9210065} ;
K.\ Li and E.\ Witten,
``{\it Role of short distance behavior in off-shell open string field
theory}'',
 \J{\PR}{D48}{853}{1993}, \hepth{9303067} ;
S.\ L.\ Shatashvili,
``{\it Comment on the background independent open string theory''},
 \J{\PL}{B311}{83}{1993}, \hepth{9303143} ;
``{\it On the problems with background independence in string theory}'',
 \hepth{9311177} ;

\bibitem{sz}
 A.\ Sen and B.\ Zwiebach,
 ``{\it Tachyon condensation in string field theory}'',
 \hepth{9912249}, \J{JHEP}{0003}{2000}{002}.

\bibitem{bpot}
 V.\ A.\ Kosteleck\'{y} and S.\ Samuel, 
 ``{\it On a Nonperturbative Vacuum for the Open Bosonic
 String}'', 
 \J{\NP}{B336}{1990}{263} ;
 N.\ Moeller and W.\ Taylor, 
 ``{\it Level truncation and the tachyon in open bosonic string field
 theory}'',
 \hepth{0002237}, \J{\NP}{B583}{2000}{105}

\bibitem{lumps}
 J.\ A.\ Harvey, and P.\ Kraus, ``{\it D-branes as unstable lumps in
 open bosonic string field theory}'',
 \hepth{0002117}, \J{JHEP}{0004}{2000}{012} ;
 R.\ de Mello Koch, A.\ Jevicki, M.\ Mihailescu and R.\ Tatar, ``{\it
 Lumps and p-branes in open string field theory}'',
 \hepth{0003031}, \J{\PL}{B482}{2000}{} ;
 N.\ Moeller,A.\ Sen and B.\ Zwiebach, ``{\it D-branes as tachyon
 lumps in string field theory}'',
 \hepth{0005036}, \J{\PL}{B482}{2000}{249} ;
 R.\ de Mello Koch and J.\ P.\ Rodrigues,
 ``{\it Lumps in level truncated open string field theory}'',
 \hepth{0008053}, \J{\PL}{B495}{2000}{237};
 N.\ Moeller,
 ``{\it Codimension two lump solutions in string field theory and
 tachyonic theories}'', \hepth{0008101}

\bibitem{nophys}
H.\ Hata and S.\ Teraguchi, 
``{\it Test of the Absence of Kinetic Terms 
around the Tachyon Vacuum in Cubic String Field Theory}'',
\hepth{0101162} ;
I.\ Ellwood and W.\ Taylor, 
``{\it Open string field theory without open strings}'',
\hepth{0103085} ;

\bibitem{vcsft}
 L.\ Rastelli, A.\ Sen, and B.\ Zwiebach, ``{\it String Field
 Theory Around the Tachyon Vacuum}'',
 \hepth{0012251}

\bibitem{vlumpsetc}
L.\ Rastelli, A.\ Sen and B.\ Zwiebach,
``{\it Classical solutions in string field theory around the 
tachyon vacuum}'',
\hepth{0102112} ;
B.\ Feng, Y.-H.\ He and N.\ Moeller, 
``{\it Testing the Uniqueness of the Open 
Bosonic String Field Theory Vacuum}'', 
\hepth{0103103} ;
I.\ Ellwood, B.\ Feng, Y.\ -H.\ He and N.\ Moeller,
``{\it The identity string field and the tachyon vacuum}'',
\hepth{0105024} ;
``{\it Half-strings, projectors, and multiple D-branes 
in vacuum string field theory}'',
\hepth{0105058} ;
D.\ J.\ Gross and W.\ Taylor,
``{\it Splitting string field theory I}
\hepth{0105059} ;
T.\ Kawano and K.\ Okuyama,
``{\it Open string fields as matrices}'',
\hepth{0105129} ;
J.\ David,
``{\it Excitations on wedge states and on the sliver}'',
\hepth{0105184}

\bibitem{rsz}
L.\ Rastelli, A.\ Sen and B.\ Zwiebach,
``{\it Boundary CFT construction of D-branes in 
vacuum string field theory}'',
\hepth{0105168}

\bibitem{bsft}
A.\ A.\ Gerasimov and S.\ L.\ Shatashvili,
``{\it On exact tachyon potential in open string field theory}'',
\J{JHEP}{0010}{2000}{034}, \hepth{0009103} ;
D.\ Kutasov, M.\ Mari\~{n}o and G.\ Moore,
``{\it Some exact results on tachyon condensation in string field theory}'',
\J{JHEP}{0010}{2000}{045}, \hepth{0009148}

\bibitem{super}
 N.\ Berkovits,
 ``{\it The Tachyon Potential in Open Neveu-Schwarz String Field
 Theory}'', 
 \hepth{0001084}, \J{JHEP}{0004}{2000}{022} ;
 N.\ Berkovits, A.\ Sen and B.\ Zwiebach,
 ``{\it Tachyon Condensation in Superstring Field Theory}'', 
 \hepth{0002211}, \J{\NP}{B587}{2000}{147} ;
 A.\ Iqbal and A.\ Naqvi,
 ``{\it Tachyon condensation on a non-BPS D-brane}'', 
 \hepth{0004015} ;
 P-J.\ de Smet and J.\ Raeymaekers,
 ``{\it Level four approximation to the tachyon potential in 
 superstring field theory}'', 
 \hepth{0003220}, \J{JHEP}{0005}{2000}{051} ;
 ``{\it The Tachyon Potential in Witten's Superstring Field Theory}'',
 \hepth{0004112}, \J{JHEP}{0008}{2000}{020} ;
 I.\ Ya.\ Aref'eva, D.\ M.\ Belov, A.\ S.\ Koshelev and P.\ B.\
 Medvedev,
 ``{\it Tachyon condensation in cubic superstring field theory}'',
 \hepth{0011117} ;
 K.\ Ohmori,
 ``{\it Tachyonic kink and lump-like solutions in superstring field 
 theory}'',
 \hepth{0104230}

\bibitem{hsms}
 H.\ Hata and S.\ Shinohara, ``{\it BRST Invariance of the
  Non-Perturbative Vacuum in Bosonic Open String Field Theory}'',
 \hepth{0009105}, \J{JHEP}{0009}{2000}{035} ;
 P.\ Mukhopadhyay and A.\ Sen, ``{\it Test of Siegel Gauge for the
 Lump Solution}'', \hepth{0101014}, \J{JHEP}{0102}{2001}{017}

\bibitem{superb}
D.\ Kutasov, M.\ Mari\~{n}o and G.\ Moore,
``{\it Remarks on tachyon condensation in superstring field theory}'',
 \hepth{0010108} ;
S.\ Moriyama and S.\ Nakamura,
``{\it Descent Relation of Tachyon Condensation from 
Boundary String Field Theory}'', 
\J{\PL}{B506}{2001}{16}, \hepth{0011002} ;
M.\ Mari\~{n}o, 
``{\it On the BV formulation of boundary superstring 
field theory}'', \hepth{0103089} ; 
V.\ Niarchos and N.\ Prezas, 
``{\it Boundary Superstring Field Theory}'',
\hepth{0103102} ;
T.\ Takayanagi, S.\ Terashima and T.\ Uesugi, 
``{\it Brane-Antibrane Action from Boundary String Field Theory}'',
 \J{JHEP}{0103}{2001}{019}, \hepth{0012210} ;
P.\ Kraus and F.\ Larsen, ``{\it Boundary String Field Theory of the 
 $D\overline{D}$ System}'', \hepth{0012198} ;
S.\ Frolov,
``{\it On off-shell structure of open string sigma model}'',
\hepth{0104042} ;
O.\ Andreev,
``{\it More about partition function of open bosonic string in background 
fields and string theory effective action}'',
\hepth{0104061} ;
K.\ S.\ Viswanathan and Y.\ Yang,
``{\it Tachyon condensation and background independent 
superstring field theory}'',
\hepth{0104099} ;
M.\ Alishahiha,
``{\it One-loop correction of the tachyon action
 in boundary superstring field theory}'',
\hepth{0104164} ;
S.\ Nakamura,
``{\it Closed string tachyon condensation and on-shell effective action
of open string tachyons}'',
\hepth{0105054} ;
K.\ Bardakci and A.\ Konechny,
``{\it Tachyon condensation in boundary string field theory at one loop}'',
\hepth{0105098} ;
B.\ Craps, P.\ Kraus and F.\ Larsen,
``{\it Loop corrected tachyon condensation}'',
\hepth{0105227}

\bibitem{o1}
 K.\ Ohmori, ``{\it A Review on Tachyon Condensation in Open String 
 Field Theories}'',
 \hepth{0102085}

\bibitem{s2}
 A.\ Sen, ``{\it Universality of the tachyon potential}'',
 \hepth{9911116}, \J{JHEP}{9912}{1999}{027}

\bibitem{ars1}
 A.\ Yu.\ Alekseev, A.\ Recknagel and V.\ Schomerus,
 ``{\it Non-commutative World-volume Geometries: Branes on SU(2) and
 Fuzzy Spheres}'',
 \hepth{9908040}, \J{JHEP}{9909}{1999}{023}

\bibitem{ars2}
 A.\ Yu.\ Alekseev, A.\ Recknagel and V.\ Schomerus, ``{\it Brane
 Dynamics in Background Fluxes and Non-commutative Geometry}'',
 \hepth{0003187}, \J{JHEP}{0005}{2000}{010}

\bibitem{hnt}
 Y.\ Hikida, M.\ Nozaki, and T.\ Takayanagi, ``{\it Tachyon
 Condensation on Fuzzy Sphere and Noncommutative Solitons}'',
 \hepth{0008023}

\bibitem{bds}
 C.\ Bachas, M.\ Douglas and C.\ Schweigert, 
 ``{\it Flux stabilization of D-branes}'',
 \hepth{0003037}, \J{JHEP}{0005}{2000}{048}

\bibitem{as}
 A.\ Yu.\ Alekseev and V.\ Schomerus,
 ``{\it D-brane in the WZW model}'',
 \hepth{9812193}, \J{\PR}{D60}{1999}{061901}

\bibitem{c}
 J.\ L.\ Cardy,
 ``{\it Boundary conditions, fusion rules and Verlinde formula}'',
 \J{\NP}{B324}{1984}{581}

\bibitem{iio}
 N.\ Ishibashi, 
 ``{\it The boundary and crosscap states in conformal field theories}'',
 \J{\MPL}{A4}{1989}{251} ;
 N.\ Ishibashi and T.\ Onogi, ``{\it Conformal field theories on
 surfaces with boundaries and crosscaps}'',
 \J{\MPL}{A4}{1989}{161}

\bibitem{fffs}
 G.\ Felder, J.\ Fr\"{o}hlich, J.\ Fuchs and C.\ Schweigert,
 ``{\it The geometry of WZW branes}'',
 \hepth{9909030}, \J{J.\ Geom.\ Phys.}{34}{2000}{162}

\bibitem{r}
 I.\ Runkel,
 ``{\it Boundary structure constants for the A-series Virasoro 
 minimal models}'',
 \hepth{9811178}, \J{\NP}{B549}{1999}{563}

\bibitem{s3}
 A.\ Sen,
  ``{\it Open string field theory in nontrivial background field: 
Gauge invariant action}'',
 \J{\NP}{B334}{350}{1990} ;
``{\it Open string field theory in nontrivial background field. 2.
Feynman rules and four point amplitudes}'',
 \J{\NP}{B334}{395}{1990} ;
``{\it Open string field theory in nontrivial background field. 3. 
N point amplitude}'',
 \J{\NP}{B335}{435}{1990}

\bibitem{rz}
 L.\ Rastelli and B.\ Zwiebach, 
 ``{\it Tachyon potentials, star products and universality}'',
 \hepth{0006240}

\bibitem{h}
 J.\ Hoppe, 
 ``{\it Diffeomorphism groups, quantization and SU($\infty$)}'',
 \J{Int.\ J.\ Mod.\ Phys.}{A4}{1989}{5235}

\bibitem{ncsoliton}
R.\ Gopakumar, S.\ Minwalla and A.\ Strominger,
``{\it Noncommutative solitons}'',
 \hepth{0003160}, \J{JHEP}{0005}{2000}{020} ; 
K.\ Dasgupta, S.\ Mukhi and G.\ Rajesh,
``{\it Noncommutative tachyons}'',
 \hepth{0005006}, \J{JHEP}{0006}{2000}{022} ; 
J.\ A.\ Harvey, P.\ Kraus, F.\ Larsen and E.\ J.\ Martinec,
``{\it D-branes and strings as non-commutative solitons}'',
 \hepth{0005031}, \J{JHEP}{0007}{2000}{042} 

\bibitem{w2}
 E.\ Witten, 
 ``{\it Noncommutative tachyons and string field theory}'',
 \hepth{0006071}

\bibitem{cs}
 L.\ Cornalba and R.\ Schiappa, 
 ``{\it Nonassociative star product deformations for D-brane
 worldvolumes in curved backgrounds}'',
 \hepth{0101219}

\bibitem{a-ggs}
 L.\ Alvarez-Gaum\'{e}, C.\ Gomez and G.\ Sierra,
 ``{\it Quantum group interpretation of some conformal field theories}'',
 \J{\PL}{B220}{1989}{142}

\end{thebibliography}
\end{document}